\documentclass[sigconf]{acmart}
\usepackage{booktabs}
\usepackage{graphicx}
\usepackage{multirow}
\usepackage{algorithm2e}
\RestyleAlgo{ruled}

\usepackage{balance}

\AtBeginDocument{%
  \providecommand\BibTeX{{%
    \normalfont B\kern-0.5em{\scshape i\kern-0.25em b}\kern-0.8em\TeX}}}
\copyrightyear{2022}
\acmYear{2022}
\setcopyright{acmcopyright}
\acmConference[KDD '22]{Proceedings of the 28th ACM SIGKDD Conference on Knowledge Discovery and Data Mining}{August 14--18, 2022}{Washington, DC, USA}
\acmBooktitle{Proceedings of the 28th ACM SIGKDD Conference on Knowledge Discovery and Data Mining (KDD '22), August 14--18, 2022, Washington, DC, USA}
\acmPrice{15.00}
\acmDOI{10.1145/3534678.3539475}
\acmISBN{978-1-4503-9385-0/22/08}

\begin{document}
\title{HICF: Hyperbolic Informative Collaborative Filtering}
\author{Menglin Yang}
\affiliation{%
  \institution{The Chinese University of Hong Kong}
  \city{Hong Kong SAR}
  \country{China}}
\email{mlyang@cse.cuhk.edu.hk}

\author{Zhihao Li}
\affiliation{%
  \institution{Harbin Institute of Technology}
  \city{Shenzhen}
  \country{China}
}
\email{180110606@stu.hit.edu.cn}

\author{Min Zhou}
 \affiliation{
 \institution{Huawei Technologies Co.,Ltd}
 \city{Shenzhen}
 \country{China}}
 \email{zhoumin27@huawei.com}

\author{Jiahong Liu}
\affiliation{%
  \institution{Harbin Institute of Technology}
  \city{Shenzhen}
  \country{China}}
  \email{jiahong.liu21@gmail.com}

\author{Irwin King}
\affiliation{%
  \institution{The Chinese University of Hong Kong}
  \city{Hong Kong SAR}
  \country{China}}
  \email{king@cse.cuhk.edu.hk}

\renewcommand{\shortauthors}{Menglin Yang et al.}
\begin{abstract}
Considering the prevalence of the power-law distribution in user-item networks, hyperbolic space has attracted considerable attention and achieved impressive performance in the recommender system recently. The advantage of hyperbolic recommendation lies in that its exponentially increasing capacity is well-suited to describe the power-law distributed user-item network whereas the Euclidean equivalent is deficient. Nonetheless, it remains unclear which kinds of items can be effectively recommended by the hyperbolic model and which cannot.
To address the above concerns, we take the most basic recommendation technique, collaborative filtering, as a medium, to investigate the behaviors of hyperbolic and Euclidean recommendation models. The results reveal that (1) tail items get more emphasis in hyperbolic space than that in Euclidean space, but there is still ample room for improvement; (2) head items receive modest attention in hyperbolic space, which could be considerably improved; (3) and nonetheless, the hyperbolic models show more competitive performance than Euclidean models. 
Driven by the above observations, we design a novel learning method, named hyperbolic informative collaborative filtering (HICF), aiming to compensate for the recommendation effectiveness of the head item while at the same time improving the performance of the tail item. 
The main idea is to adapt the hyperbolic margin ranking learning, making its pull and push procedure geometric-aware, and providing informative guidance for the learning of both head and tail items. Extensive experiments back up the analytic findings and also show the effectiveness of the proposed method.
The work is valuable for personalized recommendations since it reveals that the hyperbolic space facilitates modeling the tail item, which often represents user-customized preferences or new products.
\end{abstract}

\maketitle  
\section{Introduction}

With the growth of Amazon, Netflix, TikTok, and other e-commerce or social networking services over the past several years, recommender systems are becoming ubiquitous in the digital age.
Recommender systems, in a broad sense, are algorithms that try to suggest relevant or potentially preferable items to the users, where items are, e.g., news to read, movies to watch, goods to buy, etc.

Collaborative filtering, one of the most extensively used  techniques in the customized recommendation, is based on the assumption that users often get the preferable suggestions from someone with similar preferences. To provide relevant recommendations, collaborative-filtering approaches~~\cite{koren2009matrix,koren2008factorization,VAECF2018,nmf-cf} rely on historical interactions between users and items, which are stored in the user-item matrix. 
Recently, researchers have proposed explicitly incorporating high-order collaborative interaction to improve recommendation performance. Usually, the user-item relationship is modeled as a bipartite graph with nodes representing users or items, and edges representing their interactions. Then, graph neural networks (GNNs)~\cite{gcn2017,GAT,graphsage,zixingcikm2021,zhang2022graph} are applied to extract high-order relationships between users and items via the message propagation paradigm. By using layers of neighborhood aggregation under the graph convolutional setup to construct the final representations, these techniques~\cite{wang2019ngcf,he2020lightgcn,sun2021hgcf,mao2021simplex,mao2021ultragcn} have attained state-of-the-art performance on diverse benchmark datasets.

\begin{figure}[!t]
\centering
\includegraphics[width=0.40\textwidth]{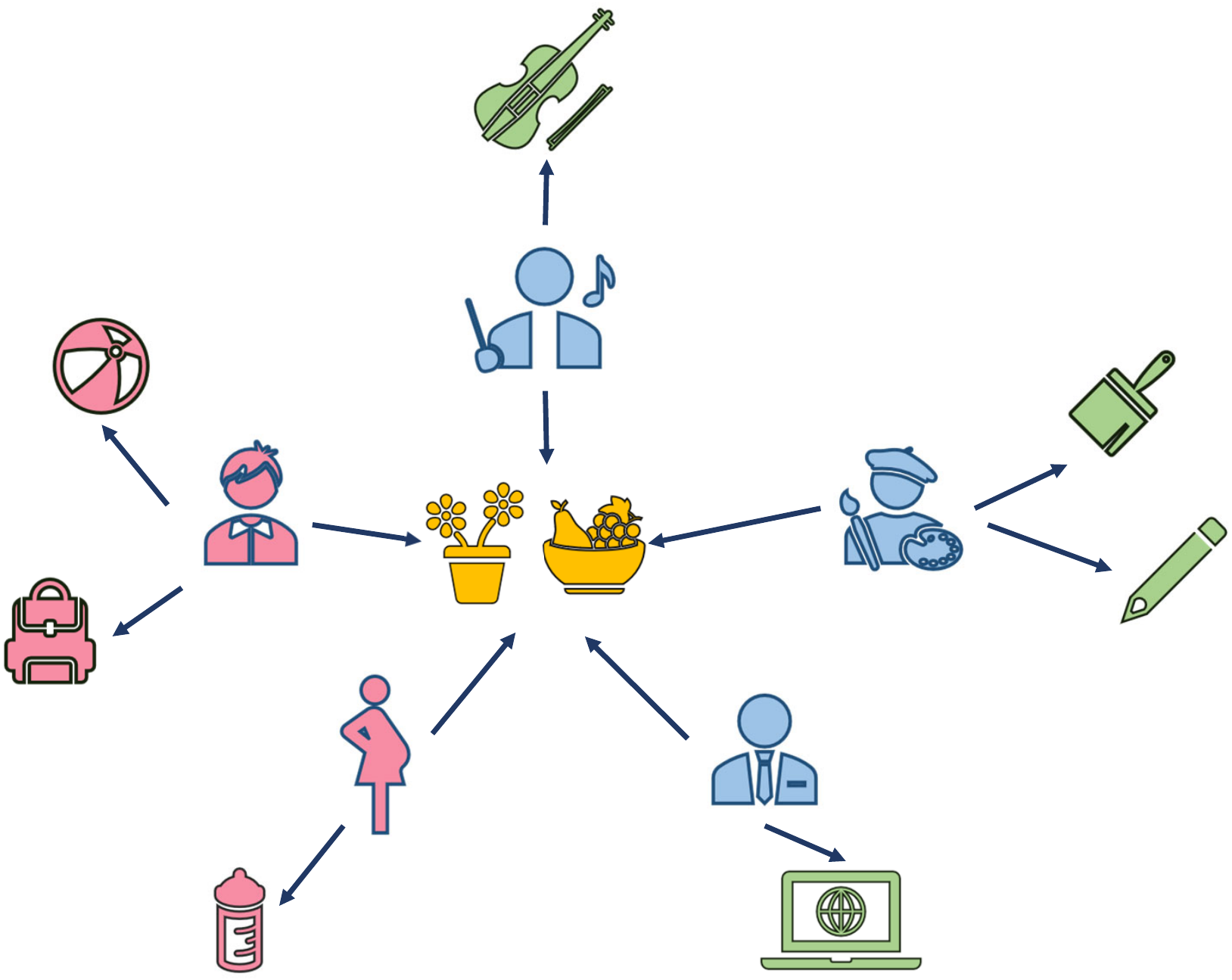}
\caption{Illustration of a user-item graph, in which the popular items, e.g., flower and fruit (the yellow icons) are liked by everyone while the preference of the other items is personalized and individualized, e.g., the paintbrush is only enjoyed by the painter and the guitar is merely appreciated by the musician. 
It is worth mentioning that the popular items are in the minority, whereas the individualized or unpopular items are in the majority.
}
\label{fig:user_item_graph}
\end{figure}

The heavy-tailed distribution\footnote{Heavy-tailed distributions are substantially right-skewed, with a small number of large values in the head and a large number of small values in the tail; they are often described by a power law, a log-normal, or an exponential function.} occurs in most large-scale recommendation datasets where the number of popular items liked by a large number of users accounts for the minority and the rest are the majority which are unpopular ones. In general, popular items are competitive while the long-tail item reflects personalized preference or something new. Both are critical for the recommendation. An example is illustrated in Figure~\ref{fig:user_item_graph}.
Recently, hyperbolic space has gained increasing interest in the recommendation area as the capacity of hyperbolic space exponentially increases with radius, which fits nicely with a power-law distributed user-item network.  Naturally, models based on hyperbolic graph neural networks achieve competitive performance in recommender systems~\cite{sun2021hgcf,chen2021modeling,yang2022hrcf}. However, it is not clear in what respects the hyperbolic model is superior to the Euclidean counterpart. At the same time, it is unclear in which aspects hyperbolic models perform worse than Euclidean models.

To answer the above doubts, in this work, we take the simplest form of recommendation model, collaborative filtering (CF), as an example to analyze and observe the behaviors of hyperbolic and Euclidean models. Specifically, we take LightGCN~\cite{he2020lightgcn} and HGCF~\cite{sun2021hgcf} as an example for analysis and observation, both of which are essentially the same model applied in different spaces. 
Specifically, we compare the recommendation effects of hyperbolic and Euclidean models, as well as their performance on the head-and-tail item, using a similar model configuration and running environment. Head and tail items are essentially chosen by the 20/80 rule\footnote{The 80/20 rule is mathematically expressed as a power-law distribution (also known as a Pareto distribution).}, which states that all items are ranked according to their degrees, and the top 20\% are considered the head (abbreviated H20), while the remaining 80\% are called the tail (abbreviated T20). The experimental findings are presented in Section~\ref{sec:observation}, which reveals the following facts. The tail item receives more consideration in the hyperbolic model than that in the Euclidean model, but there is still plenty of room for improvement, while the head item receives marginal attention in hyperbolic space, which might be substantially enhanced. Overall, the hyperbolic models outperform the Euclidean models. 
These findings are of great significance to the community of recommender systems since they help researchers better understand the advantages and disadvantages of hyperbolic models, as well as when and where to deploy them.

On the basis of the above insights, we develop a novel technique to improve hyperbolic recommender models. Two main aspects are considered: the recommendation effect of the tail item, as well as the issue of the insufficient weight placed on the head item. 
Given that most recommender systems pull the user and its interesting items in adjacent positions while pushing its uninterested items in distant areas, our method is carried out from the perspective of pull and push. The basic idea is to link the pull-and-push operations to hyperbolic geometry. Specifically, we design a hyperbolic aware margin ranking loss and hyperbolic informative aware negative sampling in such a way that both head and tail items get considerable attention. To summarize, the contribution of the proposed work is three-fold.

\begin{itemize}
    \item We initiate a quantitative investigation to study the behaviors of Euclidean and hyperbolic recommendation models, which reveals valuable findings and insights to the community. Specifically, it is observed that the hyperbolic model outperforms the Euclidean model in general and emphasizes more on the tail items, but with some sacrifice on the head items. This discovery is critical for the study of hyperbolic recommender systems.
    \item We present a hyperbolic informative collaborative filtering (HICF) method, which ensures that both head and tail items get sufficient attention via tightly coupling the embedding learning process to the hyperbolic geometry.
    \item \textcolor{black}{Extensive experiments demonstrate the effectiveness of the proposed method where the maximum recommendation effect on overall items versus all baselines up to 12.92\%, on head items against the hyperbolic model up to 12.50\%, and on tail items against hyperbolic model up to~\textbf{26.69\%}. It should be noted that the proposed method is not limited to CF-based models, but is also applicable to other hyperbolic recommendation models.}
\end{itemize}
\section{Related work}
Collaborative filtering (CF) is one of the most widely used techniques in recommender systems, in which users and items are parameterized as a matrix and the matrix parameters are learned by reconstructing historical user-item interactions. 
Earlier CF methods mapping both the ID of users and items to a joint latent factor space, so that user-item interactions are modeled as inner products in that space \cite{koren2008factorization,he2017neural,LRML2018,chen2017attentive}. The user-item interaction in the recommender system could well be represented by a bipartite graph. Recently, graph-based CF approaches~\cite{wang2019ngcf,he2020lightgcn,chen2021attentive} have made significant progress in capturing explicit relationships.
Existing graph neural networks, on the other hand, are mostly created in Euclidean space, which may understate the implicit power-law distribution of the user-item network.

Research~\cite{hgcn2019,liu2019HGNN,yang2022hyperbolic} demonstrates that the hyperbolic space is more embeddable, particularly when graph-structured data exhibit hierarchical and scale-free characteristics.
Hyperbolic representation learning has gained growing interest in the field of studying graph representation~\cite{liu2019HGNN,hgcn2019,lgcn,yang2021discrete,liu2022enhancing,yang2021hyper}. 
Due to the scale-free characteristic of the user-item network, hyperbolic geometry has also attracted a lot of attention and has been successfully applied to recommender systems~\cite{HyperML2020,wang2021hypersorec,feng2020hme,sun2021hgcf,zhang2021we,chen2021modeling,yang2022hrcf} in recent years.
For the recommender system, HyperML~\cite{HyperML2020} studies metric learning in a hyperbolic space for the representation of the user and the item. 
HGCF~\cite{sun2021hgcf} incorporates multiple layers of neighborhood aggregation using a hyperbolic GCN module to gather higher-order information in user-item interactions.
HSCML~\cite{zhang2021we} provides a thorough study of network embedding techniques for recommender systems. 
LKGR~\cite{chen2021modeling} attempts to learn embeddings in a hyperbolic space for the knowledge-graph-based recommender system.
The majority of the above works attempt to extend current Euclidean models to hyperbolic space. While the inherent benefits and shortage of hyperbolic space are seldom investigated. Although HSCML~\cite{HyperML2020} provides several interesting observations, they are insufficient to fully understand the hyperbolic recommender system.
\begin{figure*}[!t]
\centering
\includegraphics[width=4.1001cm]{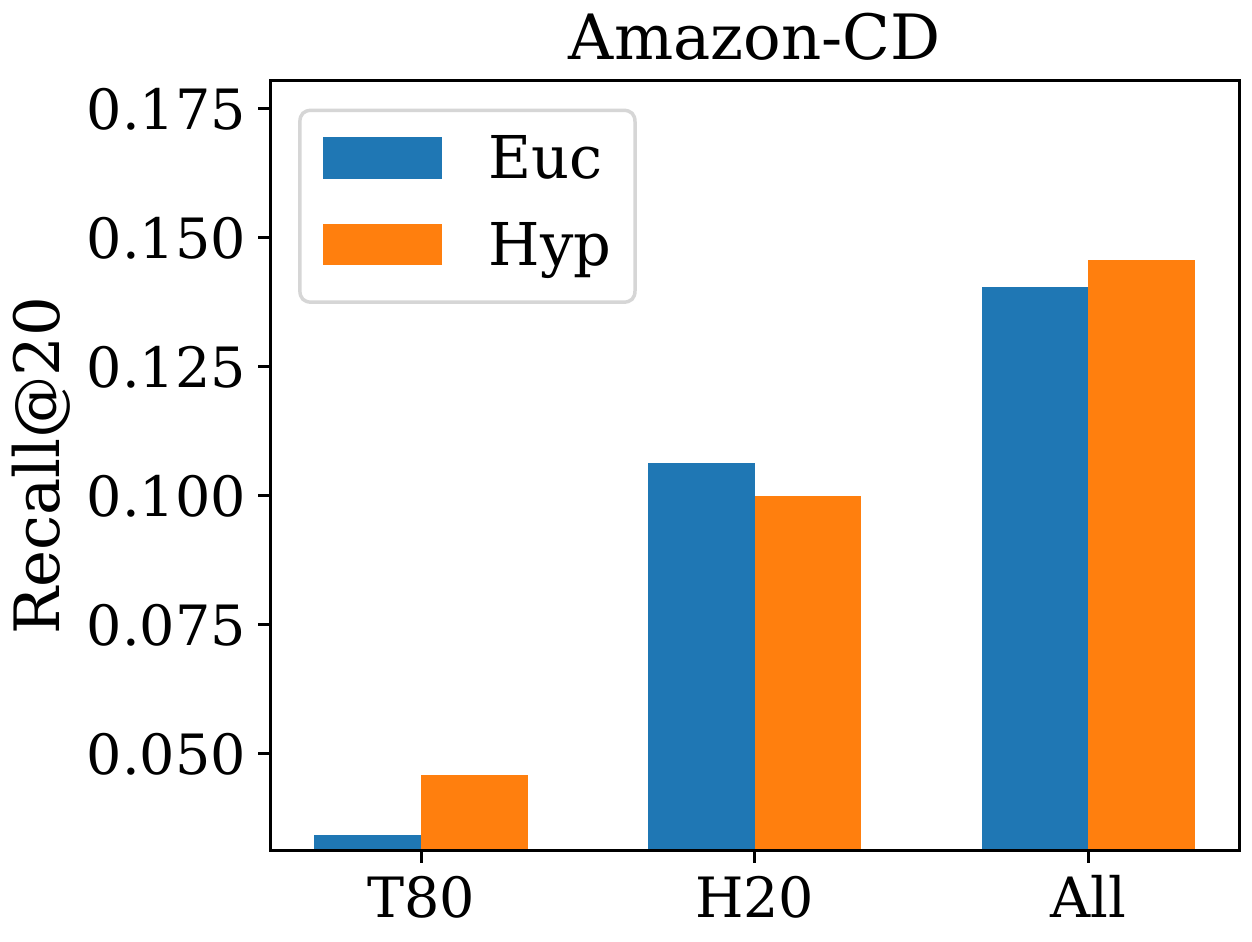}
\includegraphics[width=4.1001cm]{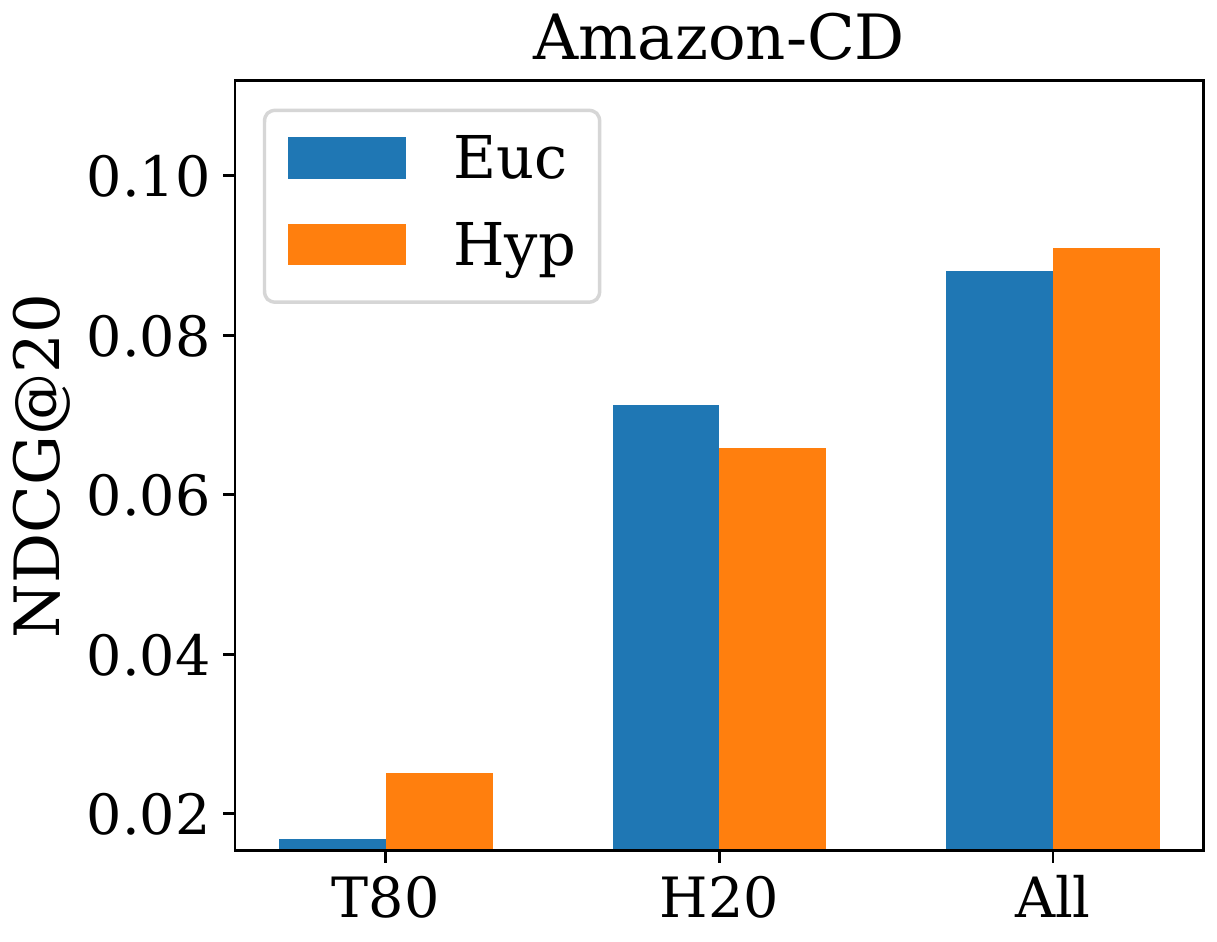}
\includegraphics[width=4.1001cm]{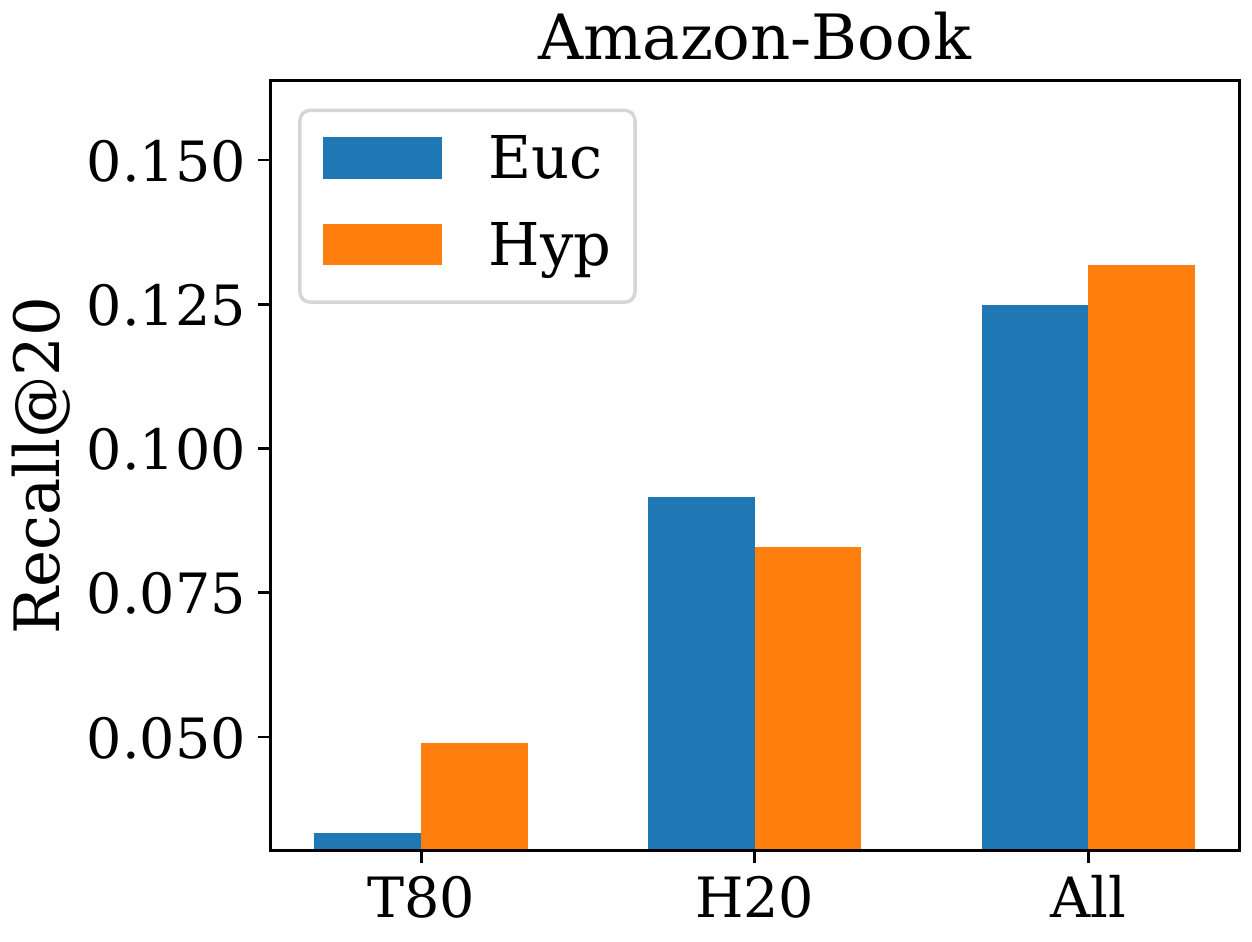}
\includegraphics[width=4.1001cm]{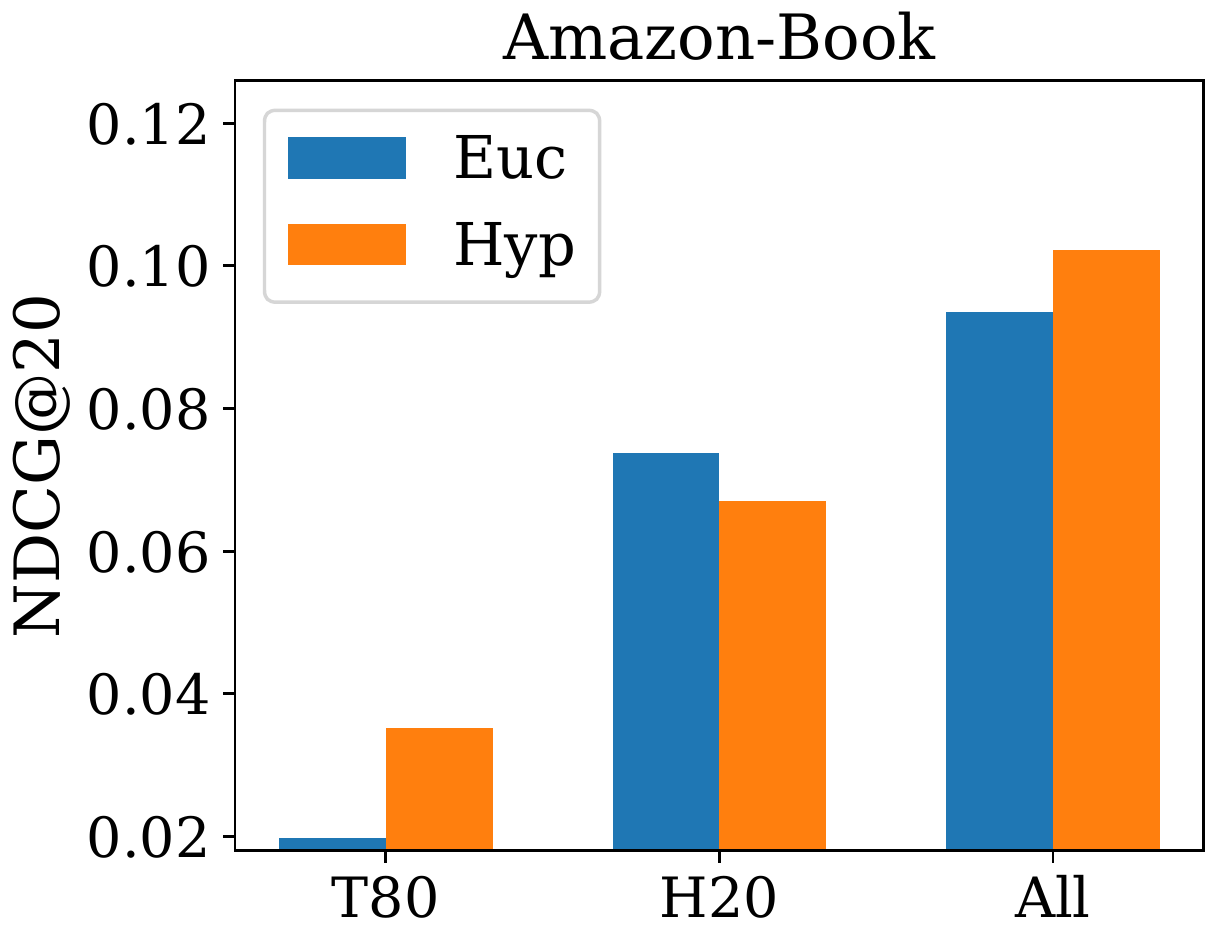}

\includegraphics[width=4.1001cm]{file/figures/top_bottom/Amazon-CD_Recall_20.pdf}
\includegraphics[width=4.1001cm]{file/figures/top_bottom/Amazon-CD_NDCG_20.pdf}
\includegraphics[width=4.1001cm]{file/figures/top_bottom/Amazon-Book_Recall_20.pdf}
\includegraphics[width=4.1001cm]{file/figures/top_bottom/Amazon-Book_NDCG_20.pdf}

\caption{Comparisons of Euclidean and hyperbolic models in Amazon-CD and Amazon-Book datasets. \texttt{Euc} represents the Euclidean model, LightGCN, and \texttt{Hyp} denotes the hyperbolic model, HGCF.}
\label{fig:obversation_on_hgcf_lightgcn}
\end{figure*}
\section{PRELIMINARIES}
Riemannian geometry is a branch of differential geometry that involves the study of smooth manifolds with a Riemannian metric.
Different curvatures of Riemannian manifolds create distinct geometries: elliptic (positive curvature), Euclidean (zero curvature), and hyperbolic (negative curvature).
We will concentrate on negative curvature space, i.e., hyperbolic geometry, in this work.
There are multiple equivalent models for hyperbolic space, each with a unique set of properties, yet being mathematically identical. The Lorentz model (alternatively called the hyperboloid model) is one of the typical hyperbolic models~\cite{nickel2018learning,hgcn2019,liu2019HGNN,lgcn}. 

An $n$-dimensional Lorentz manifold with negative curvature $-1/\kappa (\kappa>0)$ is defined as the Riemannian manifold $(\mathbb{H}^n_\kappa, g_\mathcal{L})$, where $\mathbb{H}^n_\kappa =\{\mathbf{x}\in\mathbb{R}^{n+1}:\langle\mathbf{x},\mathbf{x}\rangle_\mathcal{L}=-\kappa, x_0>0\}$, $g_\mathcal{L}=\eta$ $(\eta = \mathbf{I}_n$ except $\eta_{0,0}=-1)$ and $\langle\cdot , \cdot\rangle_\mathcal{L}$ is the Lorentzian inner product. Given $\mathbf{x,y}\in \mathbb{H}^n_\kappa$, the Lorentz inner product is give by:
\begin{equation}
    \langle\mathbf{x},\mathbf{y}\rangle_\mathcal{L}:=-x_0y_0 + \sum_{i=1}^n x_iy_i.
\label{equ:inner_product}
\end{equation}
For any $\mathbf{x}\in \mathbb{H}_\kappa^n$, there is a tangent space $\mathcal{T}_\mathbf{x}\mathbb{H}_\kappa^n$ around $\mathbf{x}$ approximating $\mathbb{H}_\kappa^n$, which is an $n$-dimensional vector space (\textit{c.f.}, Definition~\ref{def:lorentz_tangent_space}). To realize the projection between $\mathbb{H}_\kappa^n$ and $\mathcal{T}_\mathbf{x}\mathbb{H}_\kappa^n$, we can resort to the exponential map and the logarithmic map, which are given in Definition~\ref{def:lorentz_exponential_map}. The original point $\mathbf{o}:=\{\sqrt{\kappa}, 0, \cdots, 0\} \in \mathbb{H}^n_\kappa$ is a common choice as the reference point to perform these operations.

\begin{definition}[Tangent Space] 
\label{def:lorentz_tangent_space}
The {tangent space} $\mathcal{T}_\mathbf{x}\mathbb{H}_\kappa^n$ $(\mathbf{x}\in \mathbb{H}_\kappa^n)$ is defined as the first-order approximation of $\mathbb{H}_\kappa^n$ around $\mathbf{x}$:
\begin{equation}
    \mathcal{T}_\mathbf{x}\mathbb{H}_\kappa^n:=\{\mathbf{v}\in \mathbb{R}^{n+1}: \langle\mathbf{v},\mathbf{x}\rangle_\mathcal{L} = 0\}.
\end{equation}
\end{definition}

\begin{definition}[Exponential \& Logarithmic Map]
\label{def:lorentz_exponential_map}
For $\mathbf{x}\in \mathbb{H}_\kappa^n$ and $\mathbf{v}\in\mathcal{T}_\mathbf{x}\mathbb{H}_\kappa^n$ such that $\mathbf{v} \neq \mathbf{0}$ and $\mathbf{y} \neq \mathbf{x}$, there exists a unique geodesic $\gamma:[0,1]\to\mathbb{H}_\kappa^n$ where $\gamma(0)=\mathbf{x}, \gamma^\prime(0)=\mathbf{v}$.
The exponential map $\exp_\mathbf{x}: \mathcal{T}_\mathbf{x}\mathbb{H}_\kappa^n \to \mathbb{H}_\kappa^n$ is defined as $\exp_{\mathbf{x}}(\mathbf{v})=\gamma(1)$. Mathematically, 
\begin{equation}
    \exp_{\mathbf{x}}^{\kappa}(\mathbf{v})=\cosh \left(\frac{\|\mathbf{v}\|_{\mathcal{L}}}{\sqrt{\kappa}}\right) \mathbf{x} + \sqrt{\kappa} \sinh\left(\frac{\|\mathbf{v}\|_\mathcal{L}}{\sqrt{\kappa}}\right){\frac{\mathbf{v}}{\|\mathbf{v}||_{\mathcal{L}}}},
\end{equation}
where $\|\mathbf{v}\|_\mathcal{L} = \sqrt{\langle \mathbf{v}, \mathbf{v}\rangle _\mathcal{L}}$ is the Lorentzian norm of $\mathbf{v}$.
The logarithmic map $\log_\mathbf{x}$ is the inverse of the exponential $\exp_\mathbf{x}$, which is given by
\begin{equation}
    \log_{\mathbf{x}}^{\kappa}(\mathbf{y})=d_{\mathcal{L}}^\kappa(\mathbf{x},\mathbf{y})\frac{\mathbf{y}+\frac{1}{\kappa}\langle \mathbf{x}, \mathbf{y} \rangle_\mathcal{L}\mathbf{x}}{\|\mathbf{y} + \frac{1}{\kappa}\langle \mathbf{x}, \mathbf{y} \rangle_\mathcal{L}\mathbf{x}\|},
\end{equation}
where $d_\mathcal{H}^\kappa(\cdot, \cdot)$ is the distance between two points $\mathbf{x}, \mathbf{y}\in \mathbb{H}^n_\kappa$, which is formulated as:
\begin{equation}
    d_\mathcal{H}^\kappa(\mathbf{x}, \mathbf{y}) = \sqrt{\kappa}\mbox{arcosh}(-\langle \mathbf{x}, \mathbf{y}\rangle _\mathcal{L}/\kappa).
\end{equation}
\end{definition}

For simplicity, we fix $\kappa$ and set it to 1, implying that the curvature is $-1$. We will disregard $\kappa$ in the following parts for brevity.

\section{Investigation and Method}
\subsection{Hyperbolic Graph Collaborative Filtering in Brief}
The basic concept behind Euclidean and hyperbolic graph collaborative filtering~\cite{he2020lightgcn,wang2019ngcf,sun2021hgcf,chen2021modeling} is to extract high-order dependencies between users and items via a message aggregation mechanism. By graph collaborative filtering, users who like the same items, as well as items that are liked by the same users, will be grouped together.
Hyperbolic graph collaborative filtering, similar to its Euclidean counterpart, comprises three components: (1) hyperbolic encoding layer; (2) hyperbolic neighbor aggregation; and (3) prediction layer.

\textbf{Hyperbolic encoding layer.} The purpose of the hyperbolic encoding layer is to create an initial hyperbolic embedding for users and items. 
Gaussian distribution initialization is a typical method in Euclidean space. Similarly, a hyperbolic Gaussian sampling method is applied for hyperbolic recommendation models~\cite{sun2021hgcf,chen2021modeling,yang2022hrcf}. Formally, we use $\mathbf{x}\in\mathbb{R}^n$ to represent the Euclidean state of the node (including the user and the item). Then the initial hyperbolic node state $\mathbf{e}_i^0$ and $\mathbf{e}_u^0$ can be obtained by:
\begin{equation}
\begin{aligned}
    \mathbf{e}_i^0 &= \exp_\mathbf{o}(\mathbf{z}_i^0), \quad \quad \mathbf{e}_u^0 = \exp_\mathbf{o}(\mathbf{z}_u^0) \\
    \mathbf{z}_i^0 &= (0, \mathbf{x}_i), \quad\quad \mathbf{z}_u^0 = (0, \mathbf{x}_u)
\end{aligned}
    \label{equ: initialization}
\end{equation}
where $\mathbf{x}$ is taken from multivariate Gaussian distribution. $\mathbf{z}^0 = (0, \mathbf{x})$ denotes the operation of inserting the value 0 into the zeroth coordinate of $\mathbf{x}$ so that $\mathbf{z}^0$ can always live in the tangent space of origin. The superscript $0$ in $\mathbf{e}^0$ and $\mathbf{z}^0$ indicate the initial or zeroth layer state.

\textbf{Hyperbolic neighbor aggregation.} Hyperbolic neighbor aggregation is used to extract explicit user-item interaction. The hyperbolic neighbor aggregation is computed by aggregating neighboring representations of user and item from the previous aggregation. Given the neighbors $\mathcal{N}_i$ and $\mathcal{N}_u$ of $i$ and $u$, respectively, the embedding of user $u$ and $i$ is updated using the tangent state $\mathbf{z}$ and the $k$-th ($k$>0) aggregation is given by:
\begin{equation}
    \mathbf{z}_i^{k} = \mathbf{z}_i^{k-1} + \sum_{u\in \mathcal{N}_i}\frac{1}{|\mathcal{N}_i|}\mathbf{z}_u^{k-1}, \quad\quad \mathbf{z}_u^{k} = \mathbf{z}_u^{k-1} + \sum_{i\in \mathcal{N}_u}\frac{1}{|\mathcal{N}_u|}\mathbf{z}_i^{k-1}. 
\end{equation}
where $|\mathcal{N}_u|$ and $|\mathcal{N}_i|$ are the number of one-hop neighbors of $u$ and $i$, respectively.
For high-order aggregation, sum-pooling is applied in these $k$ tangential states:
\begin{equation}
    \mathbf{z}_i = \sum_{k} \mathbf{z}_i^k, \quad\quad \mathbf{z}_u = \sum_{k} \mathbf{z}_u^k.
    \label{equ:multiple aggregation}
\end{equation}
Note that $\mathbf{z}$ is on the tangent space of origin. For the hyperbolic state, it is projected back to the hyperbolic space with the exponential map,
\begin{equation}
    \mathbf{e}_i=\exp_\mathbf{o}(\mathbf{z}_i), \quad\quad\mathbf{e}_u=\exp_\mathbf{o}(\mathbf{z}_u),
\end{equation}
where $\mathbf{e}_i$ and $\mathbf{e}_u$ represents the final hyperbolic embeddings.

\textbf{Prediction layer}.
Through hyperbolic neighbor propagation, explicitly structural information is embedded in the user and item embeddings.
To infer the preference of a user for an item, the hyperbolic distance $d_\mathcal{H}$ can be used for the prediction,
$
    p(u,i) = 1/{d^2_\mathcal{H}(\mathbf{e}_u,\mathbf{e}_i)}.
$
Since we concerned with the rank of preferred items, the negative form can likewise be used for prediction, i.e, $p(u,i)=-{d^2_\mathcal{H}(\mathbf{e}_u, \mathbf{e}_i)}$.

\subsection{Investigation}
\label{sec:observation}
According to previous research~\cite{he2020lightgcn,sun2021hgcf}, the hyperbolic model~\cite{sun2021hgcf,yang2022hrcf} performs more competitively than that built in the Euclidean space~\cite{he2020lightgcn} using models with essentially the same structure.  
However, it is unclear in what aspects the hyperbolic model excels above its Euclidean equivalent. Simultaneously, it is uncertain in which places hyperbolic models are worse than Euclidean models. These issues obstruct our understanding of hyperbolic recommendation models and hinder their applications in real-world scenarios.

To solve the aforementioned doubts, we undertake a quantitative analysis that aims to experimentally study the behaviors of hyperbolic and Euclidean recommendation models by disentangling their performance on the tail and head items.  In particular, we first sort the items by their degree, which is similar to popularity, and then split them into head 20\% (denoted as {H20}, or $\mathcal{I}_{H20}$) and tail 80\%, (denoted T80, or $\mathcal{I}_{T80}$).  
Next, we investigate the effect of recommendation via the Recall@K and NDCG@K metric on H20 and T80 items, respectively, using the Euclidean graph collaborative filter model, LightGCN, and the corresponding hyperbolic model, HGCF. The results are shown in Figure~\ref{fig:obversation_on_hgcf_lightgcn}. From the experimental results, we have the following observations:
\begin{itemize}
    \item The overall recommendation performance of the hyperbolic model is better than that of the Euclidean model;
    \item Tail items get greater emphasis in the hyperbolic model as the results on tail items are far beyond that of the Euclidean counterpart ;
    \item Head items receive moderate attention in the hyperbolic model as the performance of HGCF is sightly lower than that of LightGCN.
\end{itemize}
The above results are closely related to the geometric properties of hyperbolic space:  the exponentially increased capacity of hyperbolic space enables the hyperbolic model to pay more attention to tail items compared with the Euclidean models and thus obtain an impressive performance.
Then, it is easy to know that hyperbolic recommendation models are beneficial for personalized recommendations and increasing market diversity.\footnote{As we know, the head item is popular and liked by a large number of users while the tail item is either personalized reflecting the unique preference of the user, or something fresh increasing the diversity of the market.} The hyperbolic model is a strong contender, but there are still two main shortages in the current hyperbolic model. 
(1) Despite the fact that the hyperbolic model produces better overall outcomes and has a greater recommendation effect on tail items, there is still large room for improvement. The reason is that 
tail items account for more user interests in Amazon-CD (54\% T80 vs 46\% H20) and Amazon-Book (53\% T80 vs 47\%) as given in Table~\ref{tab:datasets}, but the recommendation effect of the tail item is much lower than that of the head items, as shown in Figure~\ref{fig:obversation_on_hgcf_lightgcn}.
(2) Besides, compared with Euclidean space, hyperbolic space reduces the attention of the model on head items to a certain extent. Thus, there is an urgent need to improve the recommendation ability of head items. In this work, we aim to alleviate the above problems by improving both the head and tail items.
\subsection{Hyperbolic Informative Collaborative Filtering}

\begin{figure}[!t]
\centering
\includegraphics[width=0.45\textwidth]{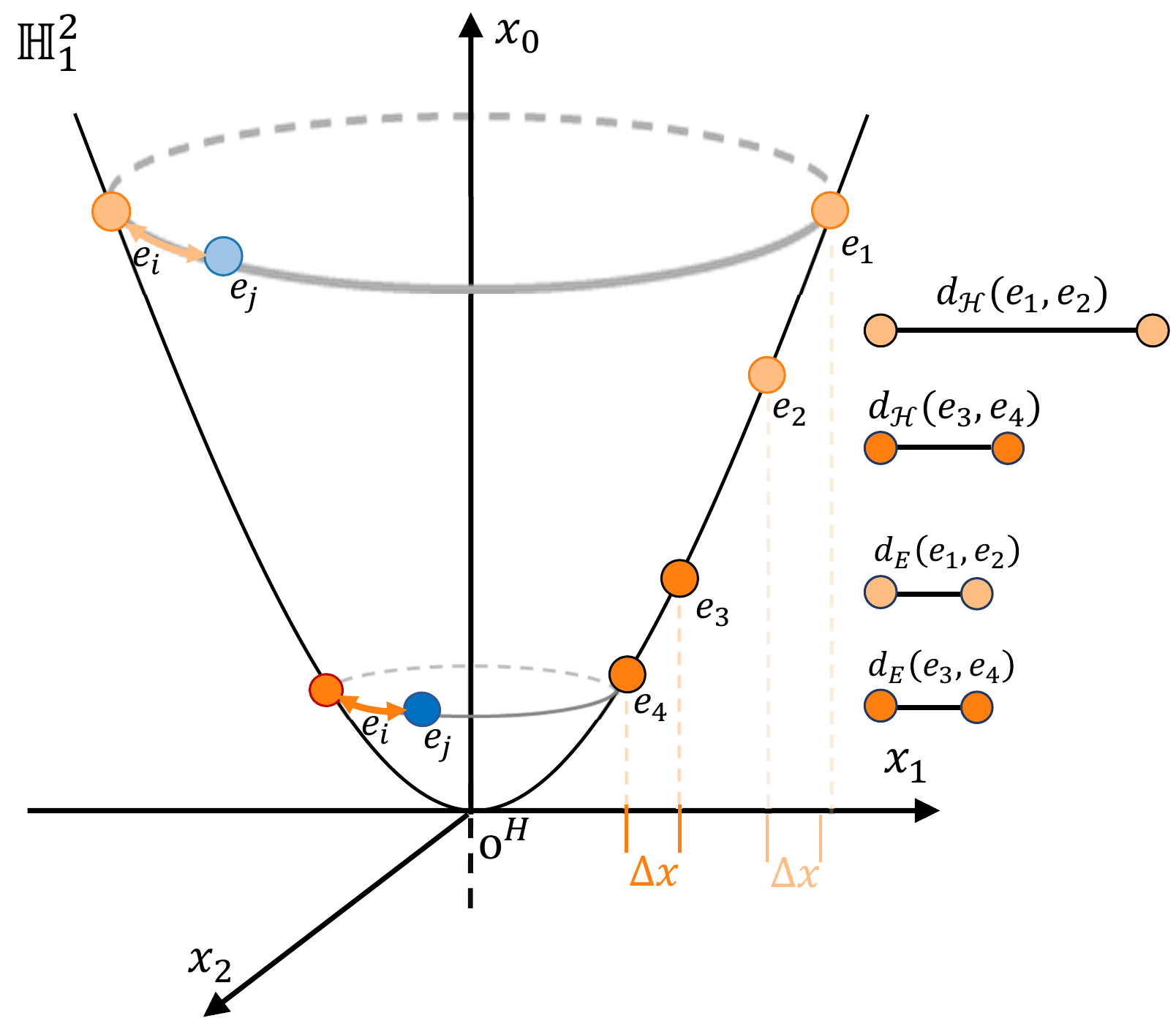}
\caption{Lorentz model of hyperbolic space $\mathbb{H}^2$ with curvature -1. On the right side of the axis $x_0$, there are two pair nodes $\{(\mathbf{e}_1, \mathbf{e}_2), (\mathbf{e}_3,\mathbf{e}_4)\}$ and their distances are equal in the Euclidean space, i.e., $d_E(\mathbf{e}_1, \mathbf{e}_2)=d_E(\mathbf{e}_3, \mathbf{e}_4)$, but are different in the hyperbolic space. In particular, the distance of the nodes $(\mathbf{e}_3, \mathbf{e}_4)$ located near the origin of hyperbolic space (HO), is smaller than the distance of $(\mathbf{e}_1, \mathbf{e}_2)$  far from HO, that is, $d_\mathcal{H}(\mathbf{e}_3, \mathbf{e}_4)<d_\mathcal{H}(\mathbf{e}_1, \mathbf{e}_2)$. On the left side of $x_0$, $\mathbf{e}_i$ and $\mathbf{e}_j$ is the positive item and the negative item, respectively, which indicates that sampling a negative item $\mathbf{e}_j$ is informative for the nodes in positions close to or far from the origin when sampling is close to the positive item $\mathbf{e}_i$.}
\label{fig:lorentz_model_of_hyperbolic_space}
\end{figure}

As we know, the optimization objectives for the representation of the user-item are generally to pull the embedding position between the user $u$ and the positive item $i$, and to push the embedding between the user $u$ and a negative item $j$. 
{The exponential growth volume} of the hyperbolic space allows samples in the hyperbolic space to be substantially more concentrated while retaining the necessary separation. 

As a result, hyperbolic margin ranking learning (HMRL)~\cite{sun2021hgcf} becomes a potent optimization tool adequately separating the items and avoiding the undesirable collapse that occurs in Euclidean space~\cite{SML2020}. HMRL, which contains two crucial components, pull and push, is used to minimize the following loss function:
\begin{equation}
    \ell(u,i,j) = \max(\underbrace{d_\mathcal{H}^2(\mathbf{e}_u, \mathbf{e}_i)}_{\text{Pull}}-\underbrace{d_\mathcal{H}^2(\mathbf{e}_u, \mathbf{e}_j)}_{\text{Push}}+m, 0), 
    \label{equ:origin_ranking_loss}
\end{equation}
where $d_\mathcal{H}$ is the hyperbolic distance and $m$ determines the margin between the distance difference between $(u, i)$ and $(u,j)$. 
Inspired by Equation~(\ref{equ:origin_ranking_loss}), we attempt to simultaneously improve the performance of the recommendation on head and tail items from the perspective of pulling and pushing. Specifically, hyperbolic informative collaborative filtering (HICF) is proposed to enable the pull and push processes to be geometric-aware.

\subsubsection{Pull: Hyperbolic-aware Margin Learning (HAML)}
Hyperbolic space is negatively curved space, meaning that as the radius increases, the volume of space grows exponentially. Put it another way, the area close to the origin is flatter, whereas the region further away curves more and has a larger capacity. As shown in Figure~\ref{fig:lorentz_model_of_hyperbolic_space}, on the right side of the axis $x_0$, there are two pair nodes $\{(\mathbf{e}_1, \mathbf{e}_2), (\mathbf{e}_3,\mathbf{e}_4)\}$ and their distances are equal in the Euclidean space, i.e. $d_E(\mathbf{e}_1, \mathbf{e}_2)=d_E(\mathbf{e}_3, \mathbf{e}_4)$, but different in the hyperbolic space. In particular, the distance of the node pair located near the \underline{o}rigin of \underline{h}yperbolic space (HO), is smaller than that far from HO, i.e., $d_\mathcal{H}(\mathbf{e}_3, \mathbf{e}_4)<d_\mathcal{H}(\mathbf{e}_1, \mathbf{e}_2)$. Such geometric properties make it natural for us to design a hyperbolic geometry-aware optimization scheme, instead of simply migrating the loss function of Euclidean space to hyperbolic space like Equation (\ref{equ:origin_ranking_loss}). 

Our main idea is to design a geometric aware margin ranking learning which enjoys the properties of hyperbolic space. Specifically, we suggest assigning a larger margin to the case where the pair (u, i) of the user and the positive item is closer to the hyperbolic origin and a smaller margin to that far away from the HO. 
It is motivated by the fact that the area close to HO is relatively flatter and the node pair is positioned in a narrow region, so a larger margin is required to SQUEEZE or PULL the pair (u, i) together, while the area distant from HO bends more and the node pair is placed in a spacious area, so a lesser margin is sufficient.\footnote{Otherwise, with the head items or high-level nodes being self-optimized close to HO referred to~\cite{nickel2017poincare,nickel2018learning,hgcn2019}, HRML, equipped with a geometric unconscious margin (e.g., a constant value), may make these head nodes difficult to be properly distinguished. This explains why the HGCF generally performs poorly on the head items.}
Inspired by the analysis (Section 2)~\cite{sala2018representation}, we know that the ratio of $\{{d_\mathcal{H}^2(\mathbf{e}_u, \mathbf{o})+d_\mathcal{H}^2(\mathbf{e}_i, \mathbf{o})\}/d_\mathcal{H}^2(\mathbf{e}_u, \mathbf{e}_i)}$ is getting smaller when they are moved away from HO. In other words, when the node pair approaches the boundary, their hyperbolic distance will approximate the sum of their hyperbolic norms. Based on these properties, we propose a hyperbolic distance-based manner to compute the margin $m_{ui}^\mathcal{H}$ which is given by:
\begin{equation}
\begin{aligned}
      m^\mathcal{H}_{ui} &= \mathrm{sigmoid}(\delta), \\
      \delta &= \frac{d_\mathcal{H}^2(\mathbf{e}_u, \mathbf{o})+d_\mathcal{H}^2(\mathbf{e}_i, \mathbf{o})-d_\mathcal{H}^2(\mathbf{e}_u, \mathbf{e}_i)}{\mathbf{e}_{u,0}\mathbf{e}_{i,0}},
\end{aligned}
\label{equ:hyperbolic_margin}
\end{equation}
where the denominator is for normalization and $\mathbf{e}_{u,0}>1$ and $\mathbf{e}_{i,0}>1$ denote the zeroth coordinate element in $\mathbf{e}_u$ and $\mathbf{e}_i$, respectively. 
$m_{ui}^\mathcal{H}$ is self-adjusting and geometry-aware, which can achieve the emphasise on both head and tail items. 
The following illustrates the intuitive understanding of Equation~(\ref{equ:hyperbolic_margin}). Since the sigmoid function increases monotonically, and we are interested in the equation of $\delta$. The numerator is the difference between the geodesic sum of $\mathbf{e}_u, \mathbf{e}_i$ to the hyperbolic origin and the hyperbolic distance of $\mathbf{e}_u, \mathbf{e}_i$, and the difference gradually decreases as the $u$ and $i$ locations move away from the hyperbolic origin~\cite{sala2018representation}. The denominator is for normalization, which increases steadily as it goes away from the hyperbolic origin. Totally, $m^\mathcal{H}_{ui}$ is getting smaller when $u$ and $i$ are moving far away from HO. Then $m^\mathcal{H}_{ui}$ is utilized to replace the $m$ in Equation~(\ref{equ:origin_ranking_loss}).

\begin{algorithm}[t]
\caption{HINS algorithm}
\label{alg:hins}
\SetKwProg{generate}{Function \emph{generate}}{}{end}
\textbf{Input:} Hyper parameters $n_{neg}$; Item set $\mathcal{I}$; the embedding matrix ${E}$; The index of the user $u$, its current positive item $i$ and its other positive item $\mathcal{N}_u$ in the training set.\\
\textbf{Output:} The informative item index $j$. \\
 Random sample $n_{neg}$ items $\mathcal{I}_u^{[n]}$ from $\mathcal{I}$\textbackslash $\mathcal{N}_{u}$\;
 \ForAll{item index $\bar{j}$ in $\mathcal{I}_u^{[n]}$}{
    Get the embeddings of the $\bar{j}$ from $E$, i.e., $\mathbf{e}_{\bar{j}}$\; 
    Let $j=-1, d_{\min} = +\infty$\;
    Compute the hyperbolic distance $d_{\mathcal{H}}(\mathbf{e}_{\bar{j}}, \mathbf{e}_i)$\; 
   \If{ $d_{\mathcal{H}}({\mathbf{e}_{\bar{j}}},\mathbf{e}_i)<d{_{\min}}$}{
      $d_{\min} = d_{\mathcal{H}}({\mathbf{e}_{\bar{j}}}, \mathbf{e}_i)$\;
      $j=\bar{j}$\;
  }
 }
\textbf{Return} $j$\;
\end{algorithm}
\subsubsection{Push: Hyperbolic Informative Negative Sampling (HINS)}
\label{sec:negative_sampling}
The basic idea is to create a training triplet, (a user $u$, the positive item $i$, the negative item $j$), by sampling the negative item $j$ from the similar popularity of item $i$ as shown in the left part of Figure~\ref{fig:lorentz_model_of_hyperbolic_space}.
This strategy provides more information than random sampling. It is simple to understand that a negative sample of a popular item is likely to be another COMPARABLE popular item while choosing the irrelevant one, such as an unpopular item, may have little effect on the optimization. 
On the other hand, the assumption of random sampling is a uniform distribution, which is incompatible with items distributed by the power law. 

In this work, we propose a self-optimizing, data-independent way to achieve negative sampling in hyperbolic space. The algorithm is demonstrated in Algorithm~\ref{alg:hins}. In each iteration, we randomly select $n_{neg}$ items $\mathcal{I}_{u}^{[n]}$, compute the hyperbolic distance between each item $\bar{j}$ in $\mathcal{I}_{u}^{[n]}$ and the positive sample $i$ and keep the item with the smallest value, where the smallest value indicates the adjacent position in the hyperbolic space. This strategy ensures that the sampled node is always close to the positive item, no matter where it is positioned, indicating that it can yield informative negative samples for both head and tail items. 

\begin{table}[!t]
\caption{Statistics of the experimental data.}
\label{tab:datasets}
\centering
\resizebox{0.48\textwidth}{!}{%
\begin{tabular}{@{}ccccccc@{}}
\toprule
\multirow{2}{*}{Dataset} & \multirow{2}{*}{\#User} & \multicolumn{3}{c}{\#Item} & \multirow{2}{*}{\#Interactions} & \multirow{2}{*}{Density} \\ \cmidrule(lr){3-5}
                         &                         & All    & H20(\%) & T80(\%) &                                 &                          \\ \midrule
Amazon-CD               & 22,947                  & 18,395 & 46    & 54    & 422,301                         & 0.10\%                   \\
Amazon-Book             & 52,406                  & 41,264 & 47    & 53    & 1,861,118                       & 0.09\%                   \\
Yelp2020                 & 71,135                  & 45,063 & 62    & 37    & 1,940,014                       & 0.05\%                   \\ \bottomrule
\end{tabular}%
}
\end{table}

\begin{table*}[h]
\caption{Recall (top table) and NDCG (bottom table) results for all datasets. The best performing model on each dataset and metric is highlighted in bold, and the second-best model is underlined. The presence of an asterisk indicates that the Wilcoxon signed-rank test for the difference in scores between the best and second-best models is statistically significant.}
\label{tab:overall_comparsion}
\resizebox{\textwidth}{!}{%
\begin{tabular}{@{}cc|cc|cccc|cc|ccc|cr@{}}
\toprule
\multicolumn{2}{c|}{Datasets}                            & WRMF   & VAE-CF & TransCF & CML    & LRML   & SML    & NGCF   & LightGCN & HAE    & HAVE   & HGCF   & Ours   & $\Delta$(\%) \\ \midrule
\multicolumn{1}{c|}{\multirow{2}{*}{Amazon-CD}}   & R@10 & 0.0863 & 0.0786 & 0.0518  & 0.0864 & 0.0502 & 0.0475 & 0.0758 & 0.0929   & 0.0666 & 0.0781 & \underline{0.0962} & \textbf{0.1079}* & +12.16    \\
\multicolumn{1}{c|}{}                             & R@20 & 0.1313 & 0.1155 & 0.0791  & 0.1341 & 0.0771 & 0.0734 & 0.1150  & 0.1404   & 0.0963 & 0.1147 & \underline{0.1455} & \textbf{0.1586}* & +9.00     \\ \midrule
\multicolumn{1}{c|}{\multirow{2}{*}{Amazon-Book}} & R@10 & 0.0623 & 0.0740  & 0.0407  & 0.0665 & 0.0522 & 0.0479 & 0.0658 & 0.0799   & 0.0634 & 0.0774 & \underline{0.0867} & \textbf{0.0965}* & +11.30    \\
\multicolumn{1}{c|}{}                             & R@20 & 0.0919 & 0.1066 & 0.0632  & 0.1023 & 0.0834 & 0.0768 & 0.1050  & 0.1248   & 0.0912 & 0.1125 & \underline{0.1318} & \textbf{0.1449}* & +9.94     \\ \midrule
\multicolumn{1}{c|}{\multirow{2}{*}{Yelp2020}}    & R@10 & 0.0470  & 0.0429 & 0.0247  & 0.0363 & 0.0326 & 0.0319 & 0.0458 & 0.0522   & 0.0360  & 0.0421 & \underline{0.0527} & \textbf{0.0570}* & +8.16     \\
\multicolumn{1}{c|}{}                             & R@20 & 0.0793 & 0.0706 & 0.0424  & 0.0638 & 0.0562 & 0.0544 & 0.0764 & 0.0866   & 0.0588 & 0.0691 & \underline{0.0884} & \textbf{0.0948}* & +7.24     \\ \bottomrule
\end{tabular}%
}

\vspace{10pt}
\resizebox{\textwidth}{!}{%
\begin{tabular}{@{}cc|cc|cccc|cc|ccc|cr@{}}
\toprule
\multicolumn{2}{c|}{Datasets}                            & WRMF   & VAE-CF & TransCF & CML    & LRML   & SML    & NGCF   & LightGCN & HAE    & HAVE   & HGCF   & Ours   & $\Delta$(\%) \\ \midrule
\multicolumn{1}{c|}{\multirow{2}{*}{Amazon-CD}}   & N@10 & 0.0651 & 0.0615 & 0.0396  & 0.0639 & 0.0405 & 0.0361 & 0.0591 & 0.0726   & 0.0565 & 0.0629 & \underline{0.0751} & \textbf{0.0848}*  & +12.92    \\
\multicolumn{1}{c|}{}                             & N@20 & 0.0817 & 0.0752 & 0.0488  & 0.0813 & 0.0492 & 0.0456 & 0.0718 & 0.0881   & 0.0657 & 0.0749 & \underline{0.0909} & \textbf{0.1010}* & +11.11    \\ \midrule
\multicolumn{1}{c|}{\multirow{2}{*}{Amazon-Book}} & N@10 & 0.0563 & 0.0716 & 0.0392  & 0.0624 & 0.0515 & 0.0422 & 0.0655 & 0.0780    & 0.0709 & 0.0778 & \underline{0.0869} & \textbf{0.0978}* & +12.54    \\
\multicolumn{1}{c|}{}                             & N@20 & 0.0730  & 0.0878 & 0.0474  & 0.0808 & 0.0626 & 0.0550  & 0.0791 & 0.0938   & 0.0789 & 0.0901 & \underline{0.1022} & \textbf{0.1142}* & +11.74    \\ \midrule
\multicolumn{1}{c|}{\multirow{2}{*}{Yelp2020}}    & N@10 & 0.0372 & 0.0353 & 0.0214  & 0.0310  & 0.0287 & 0.0255 & 0.0405 & 0.0461   & 0.0331 & 0.0371 & \underline{0.0458} & \textbf{0.0502}* & +9.13    \\
\multicolumn{1}{c|}{}                             & N@20 & 0.0506 & 0.0469 & 0.0277  & 0.0428 & 0.0369 & 0.0347 & 0.0513 & 0.0582   & 0.0409 & 0.0465 & \underline{0.0585} & \textbf{0.0633}* & +8.21     \\ \bottomrule
\end{tabular}
}
\end{table*}

\section{Experiments}
\subsection{Experimental Settings}
\textbf{Datasets.} 
In this work, we experiment with three publicly available datasets, namely Amazon-CD$^5$, Amazon-Book\footnote{https://jmcauley.ucsd.edu/data/amazon/} and Yelp2020\footnote{https://www.yelp.com/dataset}. Note that we only use user-item interactions to maintain consistency with the comparison models. The statistics of the dataset are in Table~\ref{tab:datasets}, where H20 and T80 denote the average ratio of the head items and tail items appearing in user's preference. They are calculated by $\frac{1}{|\mathcal{U}|}\sum_{u\in \mathcal{U}}\#\{\mathcal{N}_u\cap\mathcal{I}_{H20}\}$ and $\frac{1}{|\mathcal{U}|}\sum_{u\in \mathcal{U}}\#\{\mathcal{N}_u\cap\mathcal{I}_{B80}\}$, respectively. 
Each dataset is split into 80\% and 20\% training and test sets for training and evaluation, respectively.
In these datasets, ratings are transformed into binary preferences using a threshold $\geq4$ that resembles implicit feedback settings.

\textbf{Compared methods.} 
To fully verify the effectiveness of our method, we compare the baselines of the hyperbolic models and the Euclidean models. For the hyperbolic model, we compare with HGCF~\cite{sun2021hgcf}, HVAE, and HAE. HAE (HVAE) combines a (variational) autoencoder with hyperbolic geometry. Furthermore, we compare several recent strong Euclidean baselines, such as LightGCN~\cite{he2020lightgcn} and NGCF~\cite{wang2019ngcf}. In addition, we compare MF-based models, WRMF~\cite{wrmf2008} and VAE-CF~\cite{VAECF2018}; and metric learning-based models, TransCF~\cite{park2018collaborative}, CML~\cite{CML2017}, LRML~\cite{LRML2018}, and SML~\cite{SML2020}. For the data pre-processing and experimental settings, we closely follow previous work HGCF.
\begin{table*}[h]
\caption{Performance in the H20 and T80 items for all datasets. $\Delta_\mathcal{H}$ represents the relative improvements compared with the strong hyperbolic baseline HGCF. The bold denotes the best overall improvements among LightGCN, HGCF, and ours.}
\label{tab:H20_T80_results}
\resizebox{0.98\textwidth}{!}{%
\begin{tabular}{c|l|cc|cc|cc|cc|cc|cc}
\hline
\multicolumn{1}{c|}{\multirow{2}{*}{\textbf{Datasets}}} & \multicolumn{1}{l|}{\multirow{2}{*}{Models}} & \multicolumn{2}{c|}{R@20}        & \multicolumn{2}{c|}{R@10}        & \multicolumn{2}{c|}{R@5}          & \multicolumn{2}{c|}{N@20}         & \multicolumn{2}{c|}{N@10}         & \multicolumn{2}{c}{N@5}           \\  
\multicolumn{1}{r|}{}                                   & \multicolumn{1}{r|}{}                        & H20            & T80             & H20            & T80             & H20             & T80             & H20             & T80             & H20             & T80             & H20             & T80             \\ \midrule
\multirow{4}{*}{Amazon-CD}   & LightGCN      & 0.1062      & 0.0342     & 0.0741      & 0.0188     & 0.0493     & 0.0104     & 0.0712      & 0.0169     & 0.0608      & 0.0118     & 0.0529     & 0.0084     \\
                             & HGCF          & 0.0998      & 0.0457     & 0.0667      & 0.0295     & 0.0439     & 0.0179     & 0.0658      & 0.0251     & 0.0550      & 0.0201     & 0.0486     & 0.0157     \\
                             & \textbf{HICF(Ours)} & \textbf{0.1027}      & \textbf{0.0559}     & \textbf{0.0717}      & \textbf{0.0362}     & \textbf{0.0476}     & \textbf{0.0222}     & \textbf{0.0692}      & \textbf{0.0318}     & \textbf{0.0596}      & \textbf{0.0252}     & \textbf{0.0527}     & \textbf{0.0197}     \\
                             & $\Delta_\mathcal{H}(\%)$      & +2.91       & +22.32     & +7.50       & +22.71     & +8.43      & +24.02     & +5.17       & +26.69     & +8.36       & +25.37     & +8.44      & +25.48     \\ \midrule
\multirow{4}{*}{Amazon-Book} & LightGCN      & 0.0915      & 0.0333     & 0.0624      & 0.0175     & 0.0390     & 0.0104     & 0.0740      & 0.0198     & 0.0635      & 0.0145     & 0.0589     & 0.0104     \\
                             & HGCF          & 0.0829      & 0.0489     & 0.0550      & 0.0317     & 0.0344     & 0.0197     & 0.0670      & 0.0352     & 0.0578      & 0.0291     & 0.0539     & 0.0251     \\
                             & \textbf{HICF(Ours)} & \textbf{0.0898}      & \textbf{0.0551}     & \textbf{0.0603}      & \textbf{0.0362}     & \textbf{0.0387}     & \textbf{0.0227}     & \textbf{0.0738}      & \textbf{0.0404}     & \textbf{0.0642}      & \textbf{0.0336}     & \textbf{0.0605}     & \textbf{0.0293}     \\
                             & $\Delta_\mathcal{H}(\%)$      & +8.32       & +12.68     & +9.64       & +14.20     & +12.50     & +15.23     & +10.15      & +14.77     & +11.07      & +15.46     & +12.24     & +16.73     \\ \midrule
\multirow{3}{*}{Yelp2020}        & LightGCN      & 0.0836      & 0.0030     & 0.0512      & 0.0010     & 0.0298     & 0.0003     & 0.0567      & 0.0015     & 0.0448      & 0.0006     & 0.0380     & 0.0004     \\
                             & HGCF          & 0.0788      & \textbf{0.0096}     & 0.0473      & \textbf{0.0054}     & 0.0270     & 0.0030     & 0.0526      & \textbf{0.0059}     & 0.0417      & 0.0043     & 0.0354     & \textbf{0.0033}     \\
                             & \textbf{HICF(Ours)} & \textbf{0.0854}      & 0.0094     & \textbf{0.0518}      & 0.0052     & \textbf{0.0299}     & 0.0029     & \textbf{0.0576}      & 0.0057     & \textbf{0.0461}      & 0.0041     & \textbf{0.0395}     & 0.0032     \\
                             & $\Delta_\mathcal{H}(\%)$      & +8.38       & -2.08      & +9.51       & -3.70      & +10.74     & -3.33      & +9.51       & -3.39      & +10.55      & -4.65      & +11.58     & -3.03      \\ \bottomrule
\end{tabular}%
}
\end{table*}

\textbf{Experimental setup.}
For experimental settings, we closely follow the baseline HGCF to reduce the experiment burden and give a fair comparison. 
To be more specific, the number of training epochs is fixed at 500 and the embedding size is set at 50.
For gradient optimization, we use the Riemannian SGD~\cite{bonnabel2013stochastic} with weight decay in the range of $\{1e-4, 5e-4, 1e-3, 5e-3\}$ to learn the network parameters at learning rates $\{0.001, 0.0015, 0.002\}$. Note that RSGD is a stochastic gradient descent optimization technique that takes the geometry of the hyperbolic manifold into consideration. For the experimental settings of baselines, we refer to~\cite{sun2021hgcf}. 

\textbf{Evaluation metrics}.
We employ two standard evaluation metrics to assess the performance of the top-K recommendation and preference ranking: Recall and NDCG~\cite{ying2018graph}. We treat each observed interaction between a user and an item as a positive case and then use the HINS to match it with one unfavorable item that the user has not previously rated.

\subsection{Overall Performance}
The overall experimental results of the test set are summarized in Table~\ref{tab:overall_comparsion}, with the best results in bold, the second-best in italics, and $\Delta$ representing the relative improvement over the best baseline. In summary, the proposed method successfully outperforms all baselines in both Recall and NDCG metrics, with the highest improvement reaching 12.92\%, demonstrating its impressive effectiveness. We further illustrate some in-depth observations. First, hyperbolic models, including the proposed HICF and HGCF, are more competitive in modeling large-scale user-item networks than Euclidean models. The main reason is that, as networks expand, the distribution of power laws becomes more apparent. Thereby, the hyperbolic models are more competitive.
Furthermore, the improvement is observed to be relative to data density (\textit{c.f.}~Table~\ref{tab:datasets}). . Specifically, the improvements of the model in the data with higher density, e.g., Amazon-CD which are +12.16\% for Recall@10 and +12.92\% for NDCG@10 are generally greater than those in lower density data, e.g., yelp, which are +8.16\% for Recall@10 and +8.21\% for NDCG@10. 

\subsection{Performance on Head and Tail Items}
To further illustrate the validity of the proposal, we performed an in-depth analysis comparing the performance of the tail and head items separately. For simplicity, we focus on the two most prominent baselines, the hyperbolic HGCF model and its Euclidean counterpart LightGCN. The findings are listed in Table~\ref{tab:H20_T80_results}, where $\Delta_\mathcal{H}$ represents the improvements over the hyperbolic model, especially demonstrating the role of the proposed method against the original hyperbolic model. 
The corresponding overall performance can be computed by adding the results of H20 and T80 together, and the best overall performances among HICF, HRCF, and LightGCN are bold.

\textcolor{black}{From the experimental results and the bold notation in Table~\ref{tab:H20_T80_results}, we know that the performance of the proposed HICF consistently outperforms the baselines. In particular, we discovered that the performance of hyperbolic models including our HICF and HGCF on tail items is eye-catching, while the performance on head items is slightly inferior. Overall, the proposed HICF successfully achieves the aforementioned goal, e.g., enabling the hyperbolic model to improve the performance on both tail items and head items.  
In particular, for \textit{tail} items, we found that HICF performance is significantly improved compared to HGCF on Amazon-CD and Amazon-Book with the largest improvement up to 26.69\% and 16.73\%, respectively, and stays comparable on Yelp, which may be due to the fact that Yelp users show fewer interests in tail items, as seen from Table~\ref{tab:datasets}. For \textit{head} items, the performances of HICF comprehensively outperform that of HGCF and the improvements are up to 8.43\% on Amazon-CD, 12.50\% on Amazon-Book, and 11.58\% on Yelp. In addition, HICF narrows the performance gap of the tail items with Euclidean LightGCN on Amazon-CD and Amazon-Book and impressively surpasses the performance of LightGCN on Yelp.
}
\begin{figure}[!t]
\centering
\includegraphics[width=4.10cm]{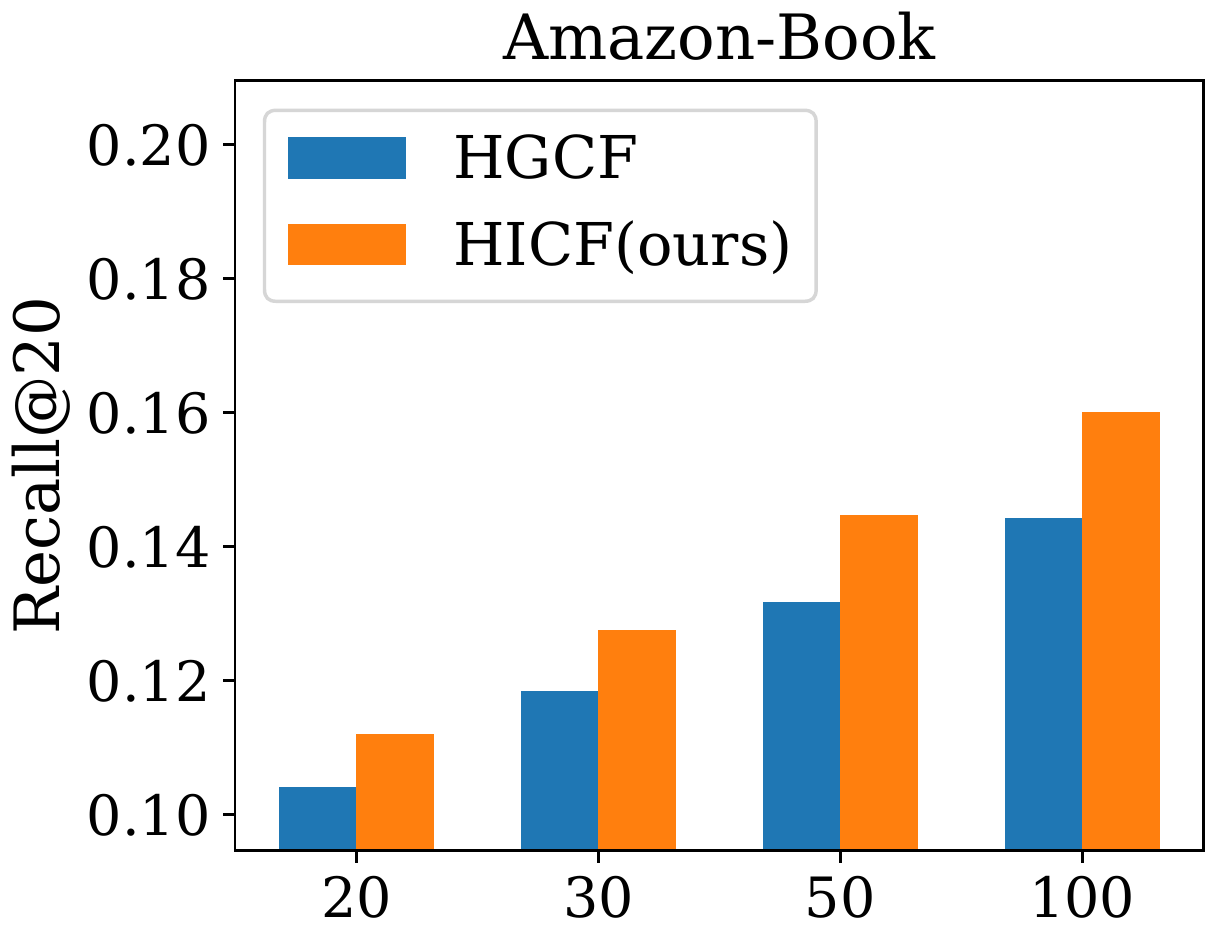}
\includegraphics[width=4.10cm]{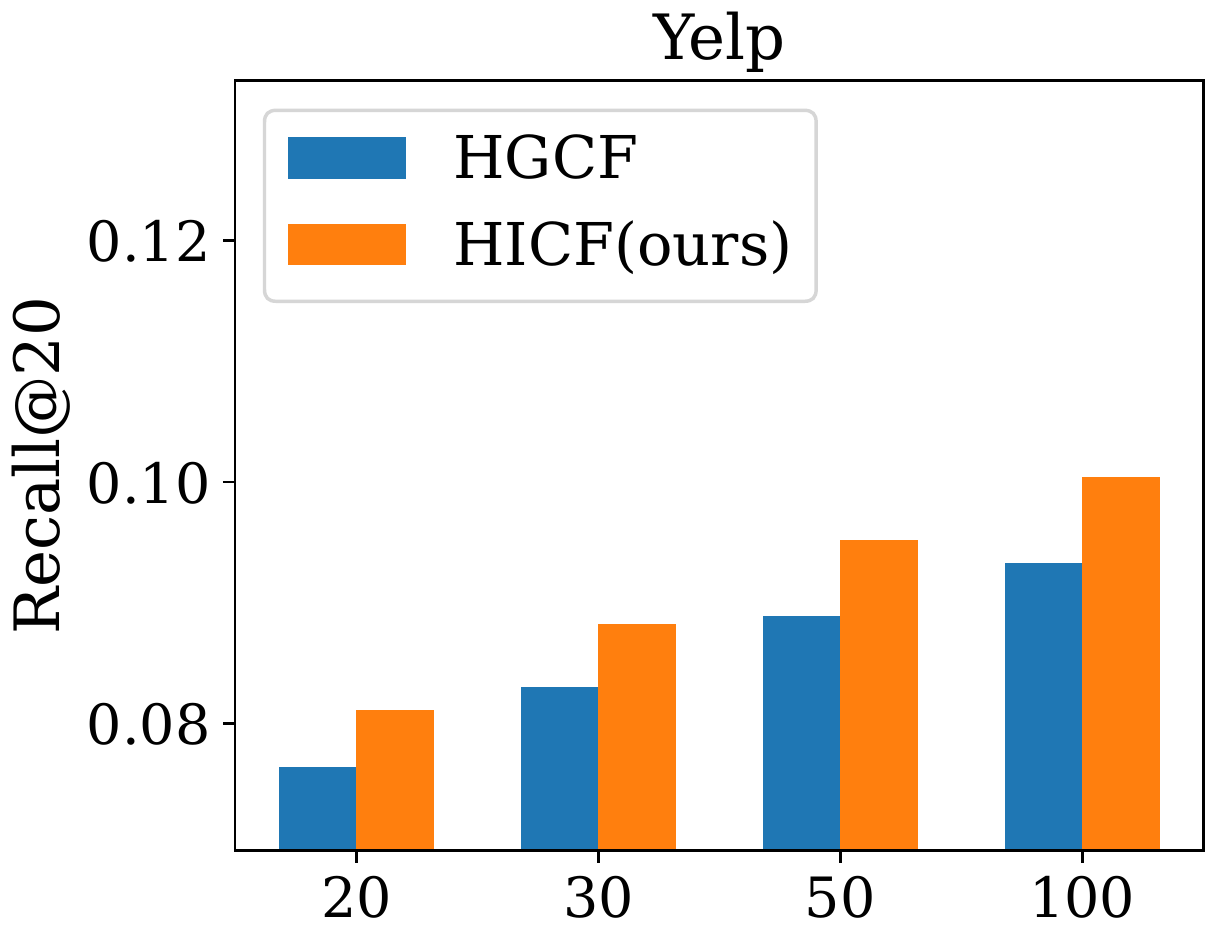}
\caption{Comparisons among four dimensions $\{20, 30, 50, 100\}$ in the Amazon-CD and Amazon-Book datasets. The evaluation metric is Recall@20, and other metrics@K show similar results. }
\label{fig:dim_study}
\end{figure}

\subsection{Generalization w.r.t. Embedding sizes}
Considering that the embedding dimension has an effect on the embedding capacity and the pairwise embedding distance, to fully verify the generalization of our proposed method, we traverse different embedding dimensions for evaluation. The experimental results are shown in Figure~\ref{fig:dim_study}. From the experimental results, we easily know that the proposed HICF continuously outperforms the strongest baseline HGCF. At the same time, we found that with the increase of the embedding dimension, the performance of the model is further improved. These findings validate the strong generalizability of the proposed HICF.

\subsection{Convergent Speed w.r.t Training Epochs}
\begin{figure}[!t]
\centering
\includegraphics[width=4.10cm]{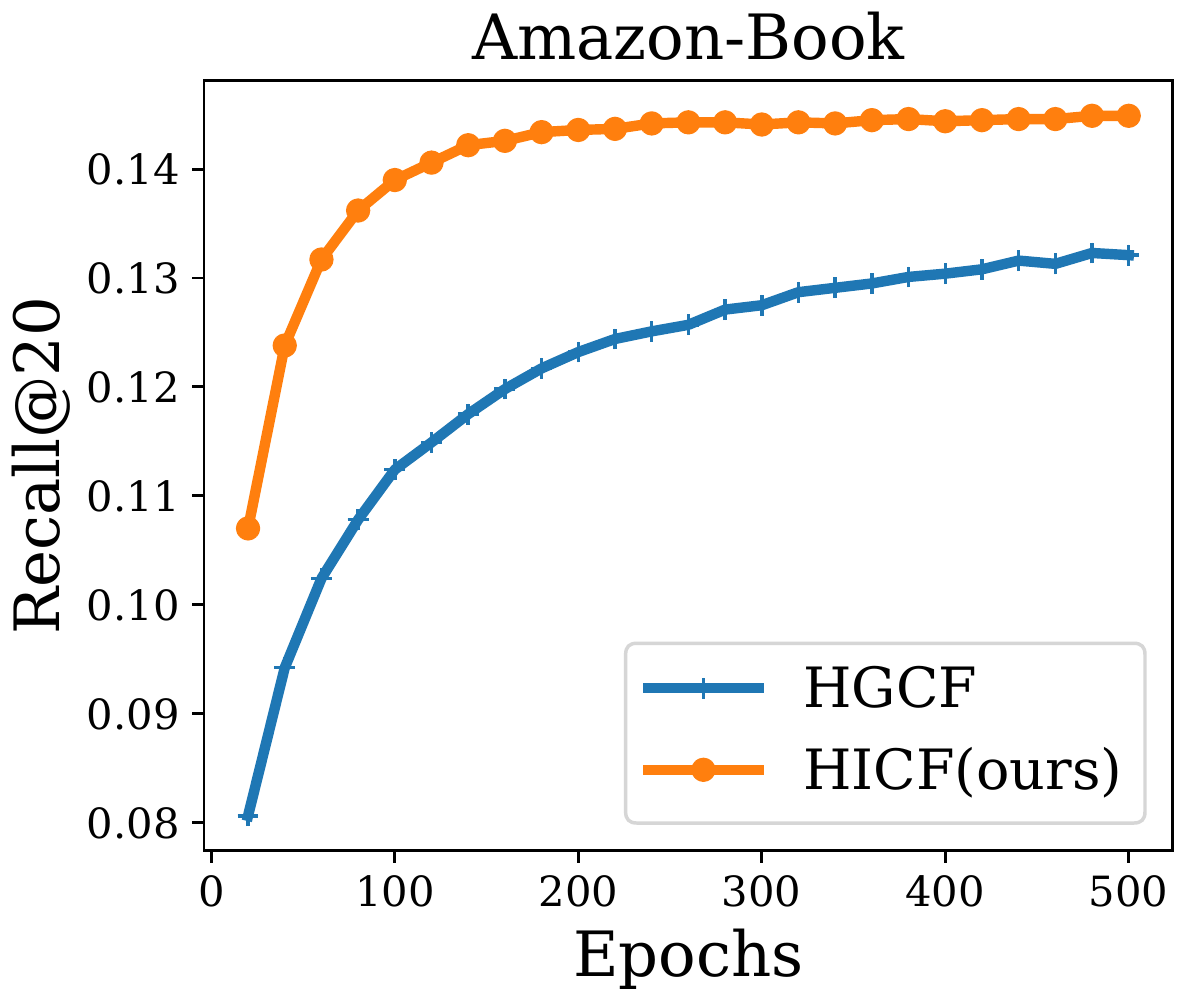}
\includegraphics[width=4.10cm]{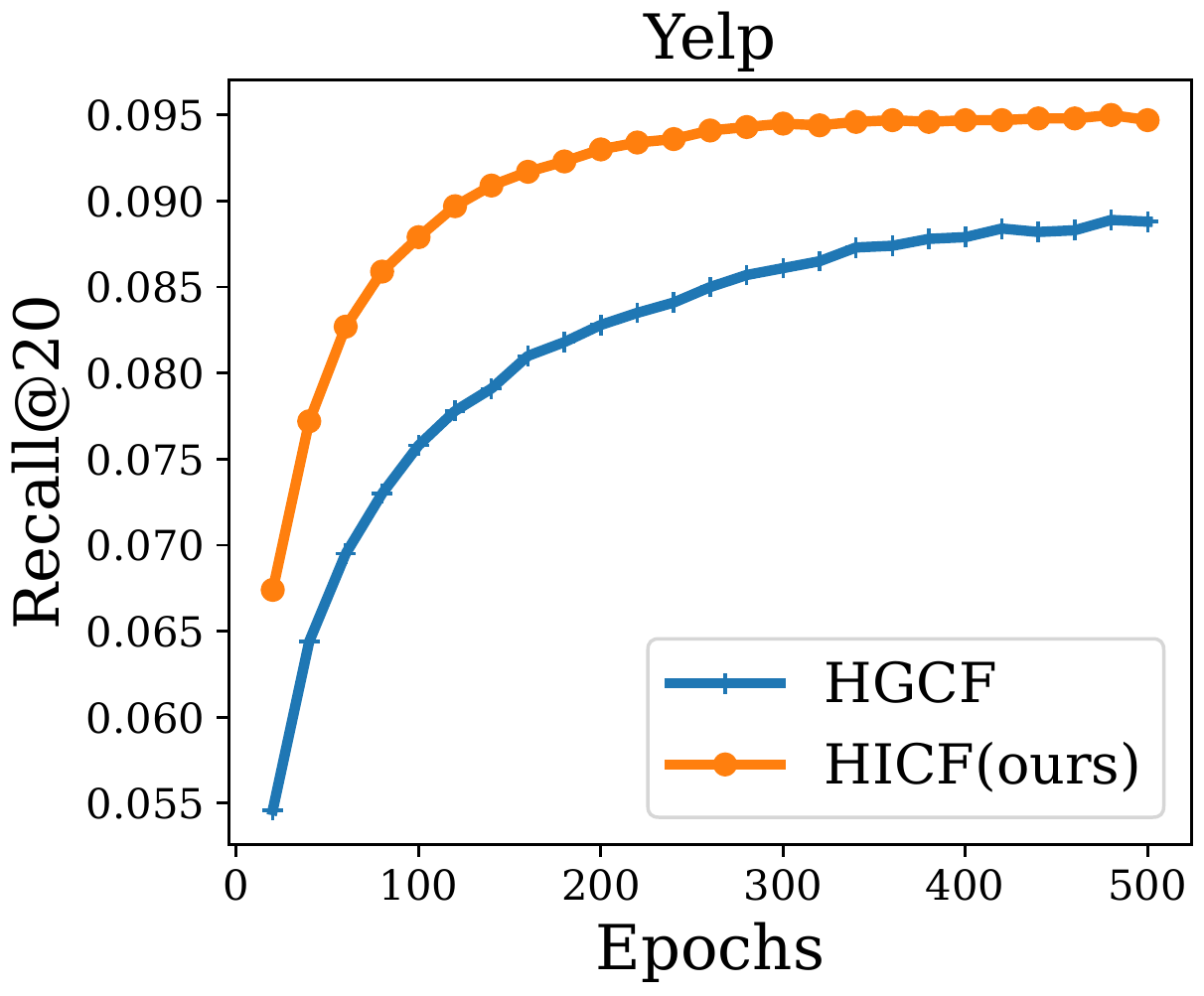}
\caption{Recall@20 changes with training epochs on Amazon-Book and Yelp. Other metric@Ks show a similar tendency.}
\label{fig:metric_training_epoch}
\end{figure}

Likewise, we analyze the convergence of the proposed approach for training the proposed HICF model and its most comparable equivalent HGCF. Figure~\ref{fig:metric_training_epoch} shows the performance of the Recall@20 metrics for epochs from 1 to 500. Other metrics@Ks show a similar tendency. The following conclusions are drawn from the experimental findings: (1) the proposed HICF repeatedly outperforms the baseline model in all epochs; (2) the proposed technique is capable of achieving the highest performance in fewer epochs, indicating that HICF can speed up the training process.

\subsection{Ablation Study}
\begin{table}[]
\caption{Ablation study (AS) for HICF. w/o M denotes without HAML and w/o S denotes without HINS in HICF.}
\centering
\label{tab:ablation_study}
\resizebox{0.48\textwidth}{!}{%
\begin{tabular}{@{}c|ccc|ccc|ccc@{}}
\toprule
\multirow{2}{*}{AS} & \multicolumn{3}{c|}{Amazon-CD} & \multicolumn{3}{c|}{Amazon-Book} & \multicolumn{3}{c}{Yelp2020} \\ \cmidrule(l){2-10} 
                  & R@20     & H20      & T80      & R@20      & H20       & T80      & R@20   & H20    & T80    \\ \midrule
HICF              & \textbf{0.1586}   & \textbf{0.1027}   & \textbf{0.0559}   & \textbf{0.1449}    & \textbf{0.0898}    & \textbf{0.0551}   & \textbf{0.0947} & \textbf{0.0854} & {0.0094} \\
w/o M             & 0.1534   & 0.0968   & 0.0566   & 0.1363    & 0.0844    & 0.0519   & 0.0901 & 0.0803 & \textbf{0.0098} \\
w/o S             & 0.1492   & 0.1021   & 0.0472   & 0.1312    & 0.0813    & 0.0499   & 0.0921 & 0.0826 & 0.0096 \\ \bottomrule
\end{tabular}%
}
\end{table}

In this part, we conduct an ablation study to evaluate the effectiveness of each component in the proposed HICF. We remove HAML and HINS separately. In Table~\ref{tab:ablation_study}, we report the corresponding results for Recall@20. Note that other metric@Ks have similar outcomes. We know that removing both HAML and HINS will lead to degradation of the model performance, which verifies the effectiveness of the proposed method. In particular, Removing HINS on both Amazon-CD and Amazon-Book leads to a large decline, while Yelp has a smaller drop, which is mainly due to the density of the dataset. The Amazon-Book and Amazon-CD dataset are relatively dense and require more negative samples for selection, while Yelp is sparse and less dependent on informative negative samples.

\subsection{Parameter Analysis}

\begin{figure}[!t]
\centering
\includegraphics[width=4.10cm]{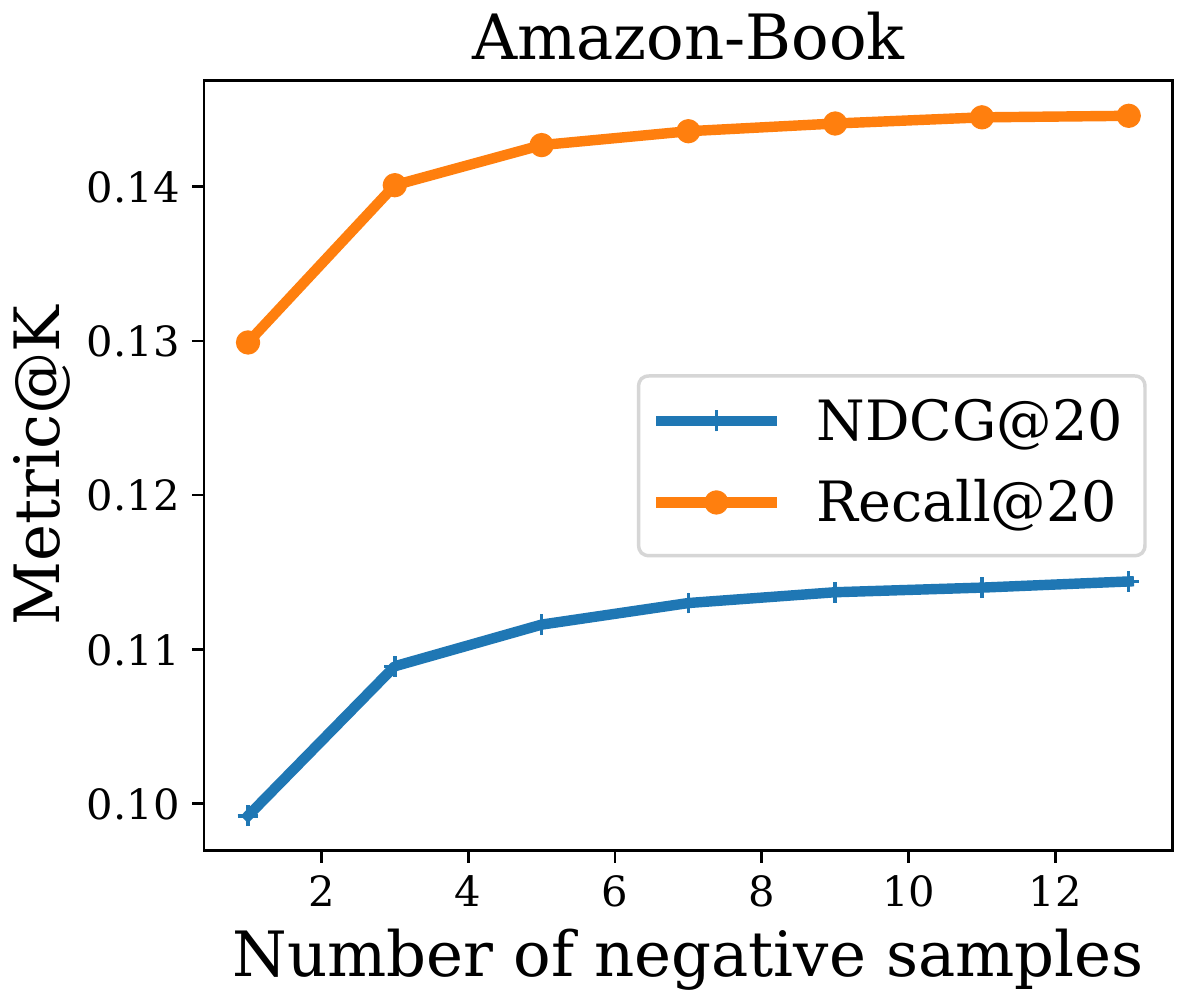}
\includegraphics[width=4.10cm]{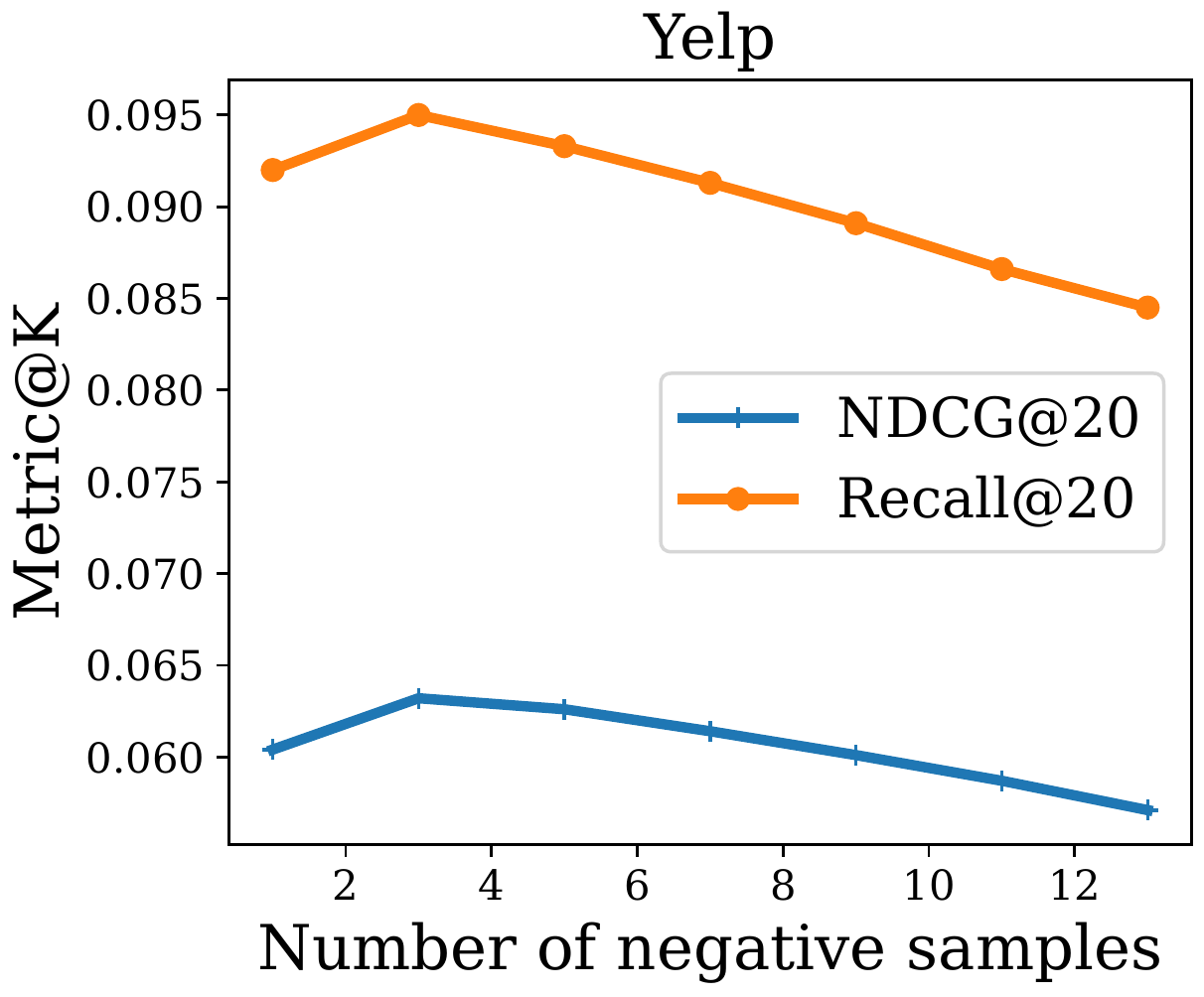}
\caption{The number of negative samples for selection. }
\label{fig:nneg_study}
\vspace{-10pt}
\end{figure}

The number of negative samples for the selection is a critical hyperparameter in this study. In our research, we discovered that it varies with each dataset, but a simple grid search enables us to quickly locate a suitable value. We do a parameter sensitivity study on Amazon-Book and Yelp in this section. As shown in Figure~\ref{fig:nneg_study}, we know that Yelp requires fewer negative samples for selection and Amazon-Book requires more negative samples for selection.
It can be understood that when the dataset (e.g., Amazon-Book) is relatively dense, each node has more neighbors, and then more negative candidates are required to find a suitable one. The influence of parameter analysis is also reflected in the ablation study of HINS.

\section{Conclusion}

Hyperbolic models have received increasing attention in the recommendation community, while their pros and cons over their Euclidean counterparts have not been explicitly studied.  In this work, we attempt to initiate the investigation by further separately comparing their performance on head-and-tail items against the Euclidean equivalents. 
Overall, the hyperbolic model shows apparent superiority. It is also observed that the hyperbolic model performs substantially better on tail items than the Euclidean equivalent, but there is still sufficient room for improvement. For the head item, the hyperbolic model gives modest attention.
Motivated by the observations, we propose a geometry boosted hyperbolic collaborative filtering (i.e., HICF). Our technique is designed to make the pull and push components of the hyperbolic margin ranking loss geometric aware, which then provides informative optimization guidance for both head and tail items. It is worth noting that our approach is not limited to collaborative filtering-based models but is also applicable to other recommendation models. 

The above observation and proposed technique shed more light on the role of hyperbolic models in the recommender system. Note that the exponentially increased capacity of hyperbolic space allows the model to pay more attention to tail items, which is beneficial for personalized recommendations and increasing market diversity. 
In future work, we aim to analyze the advantages and disadvantages of hyperbolic representation models from more generalized settings and other applications~\cite{zixingsurvey,FeatureNorm2020, li2019improving,li2020unsupervised,li2022text,li2022bsal,zhou2022telegraph}.

\begin{acks}
The work described in this paper was partially supported by the National Key Research and Development Program of China (No. 2018AAA0100204) and the Research Grants Council of the Hong Kong Special Administrative Region, China (CUHK 2410021, Research Impact Fund, No. R5034-18). We thank the anonymous reviewers for their constructive comments.
\end{acks}
\bibliographystyle{ACM-Reference-Format}
\balance 
\bibliography{reference}


\begin{thebibliography}{48}


\ifx \showCODEN    \undefined \def \showCODEN     #1{\unskip}     \fi
\ifx \showDOI      \undefined \def \showDOI       #1{#1}\fi
\ifx \showISBNx    \undefined \def \showISBNx     #1{\unskip}     \fi
\ifx \showISBNxiii \undefined \def \showISBNxiii  #1{\unskip}     \fi
\ifx \showISSN     \undefined \def \showISSN      #1{\unskip}     \fi
\ifx \showLCCN     \undefined \def \showLCCN      #1{\unskip}     \fi
\ifx \shownote     \undefined \def \shownote      #1{#1}          \fi
\ifx \showarticletitle \undefined \def \showarticletitle #1{#1}   \fi
\ifx \showURL      \undefined \def \showURL       {\relax}        \fi
\providecommand\bibfield[2]{#2}
\providecommand\bibinfo[2]{#2}
\providecommand\natexlab[1]{#1}
\providecommand\showeprint[2][]{arXiv:#2}

\bibitem[\protect\citeauthoryear{Bonnabel}{Bonnabel}{2013}]%
        {bonnabel2013stochastic}
\bibfield{author}{\bibinfo{person}{Silvere Bonnabel}.}
  \bibinfo{year}{2013}\natexlab{}.
\newblock \showarticletitle{Stochastic gradient descent on Riemannian
  manifolds}.
\newblock \bibinfo{journal}{\emph{TAC}} \bibinfo{volume}{58},
  \bibinfo{number}{9} (\bibinfo{year}{2013}), \bibinfo{pages}{2217--2229}.
\newblock


\bibitem[\protect\citeauthoryear{Chami, Ying, R{\'e}, and Leskovec}{Chami
  et~al\mbox{.}}{2019}]%
        {hgcn2019}
\bibfield{author}{\bibinfo{person}{Ines Chami}, \bibinfo{person}{Zhitao Ying},
  \bibinfo{person}{Christopher R{\'e}}, {and} \bibinfo{person}{Jure Leskovec}.}
  \bibinfo{year}{2019}\natexlab{}.
\newblock \showarticletitle{Hyperbolic graph convolutional neural networks}. In
  \bibinfo{booktitle}{\emph{NeurIPS}}. \bibinfo{pages}{4868--4879}.
\newblock


\bibitem[\protect\citeauthoryear{Chen, Zhang, He, Nie, Liu, and Chua}{Chen
  et~al\mbox{.}}{2017}]%
        {chen2017attentive}
\bibfield{author}{\bibinfo{person}{Jingyuan Chen}, \bibinfo{person}{Hanwang
  Zhang}, \bibinfo{person}{Xiangnan He}, \bibinfo{person}{Liqiang Nie},
  \bibinfo{person}{Wei Liu}, {and} \bibinfo{person}{Tat-Seng Chua}.}
  \bibinfo{year}{2017}\natexlab{}.
\newblock \showarticletitle{Attentive collaborative filtering: Multimedia
  recommendation with item-and component-level attention}. In
  \bibinfo{booktitle}{\emph{SIGIR}}. \bibinfo{pages}{335--344}.
\newblock


\bibitem[\protect\citeauthoryear{Chen, Yang, Zhang, Zhao, Meng, Hao, and
  King}{Chen et~al\mbox{.}}{2022b}]%
        {chen2021modeling}
\bibfield{author}{\bibinfo{person}{Yankai Chen}, \bibinfo{person}{Menglin
  Yang}, \bibinfo{person}{Yingxue Zhang}, \bibinfo{person}{Mengchen Zhao},
  \bibinfo{person}{Ziqiao Meng}, \bibinfo{person}{Jianye Hao}, {and}
  \bibinfo{person}{Irwin King}.} \bibinfo{year}{2022}\natexlab{b}.
\newblock \showarticletitle{Modeling Scale-free Graphs with Hyperbolic Geometry
  for Knowledge-aware Recommendation}. In \bibinfo{booktitle}{\emph{WSDM}}.
  \bibinfo{publisher}{{ACM}}, \bibinfo{pages}{94--102}.
\newblock


\bibitem[\protect\citeauthoryear{Chen, Yang, Wang, Bai, Song, and King}{Chen
  et~al\mbox{.}}{2022a}]%
        {chen2021attentive}
\bibfield{author}{\bibinfo{person}{Yankai Chen}, \bibinfo{person}{Yaming Yang},
  \bibinfo{person}{Yujing Wang}, \bibinfo{person}{Jing Bai},
  \bibinfo{person}{Xiangchen Song}, {and} \bibinfo{person}{Irwin King}.}
  \bibinfo{year}{2022}\natexlab{a}.
\newblock \showarticletitle{Attentive Knowledge-aware Graph Convolutional
  Networks with Collaborative Guidance for Personalized Recommendation}. In
  \bibinfo{booktitle}{\emph{ICDE}}.
\newblock


\bibitem[\protect\citeauthoryear{Feng, Tran, Cong, Chen, Li, and Li}{Feng
  et~al\mbox{.}}{2020}]%
        {feng2020hme}
\bibfield{author}{\bibinfo{person}{Shanshan Feng}, \bibinfo{person}{Lucas~Vinh
  Tran}, \bibinfo{person}{Gao Cong}, \bibinfo{person}{Lisi Chen},
  \bibinfo{person}{Jing Li}, {and} \bibinfo{person}{Fan Li}.}
  \bibinfo{year}{2020}\natexlab{}.
\newblock \showarticletitle{{HME}: A hyperbolic metric embedding approach for
  next-poi recommendation}. In \bibinfo{booktitle}{\emph{SIGIR}}.
  \bibinfo{pages}{1429--1438}.
\newblock


\bibitem[\protect\citeauthoryear{Hamilton, Ying, and Leskovec}{Hamilton
  et~al\mbox{.}}{2017}]%
        {graphsage}
\bibfield{author}{\bibinfo{person}{William~L Hamilton}, \bibinfo{person}{Rex
  Ying}, {and} \bibinfo{person}{Jure Leskovec}.}
  \bibinfo{year}{2017}\natexlab{}.
\newblock \showarticletitle{Inductive representation learning on large graphs}.
  In \bibinfo{booktitle}{\emph{ICONIP}}. \bibinfo{pages}{1025--1035}.
\newblock


\bibitem[\protect\citeauthoryear{He, Deng, Wang, Li, Zhang, and Wang}{He
  et~al\mbox{.}}{2020}]%
        {he2020lightgcn}
\bibfield{author}{\bibinfo{person}{Xiangnan He}, \bibinfo{person}{Kuan Deng},
  \bibinfo{person}{Xiang Wang}, \bibinfo{person}{Yan Li},
  \bibinfo{person}{Yongdong Zhang}, {and} \bibinfo{person}{Meng Wang}.}
  \bibinfo{year}{2020}\natexlab{}.
\newblock \showarticletitle{Light{GCN}: Simplifying and powering graph
  convolution network for recommendation}. In
  \bibinfo{booktitle}{\emph{SIGIR}}. \bibinfo{pages}{639--648}.
\newblock


\bibitem[\protect\citeauthoryear{He, Liao, Zhang, Nie, Hu, and Chua}{He
  et~al\mbox{.}}{2017}]%
        {he2017neural}
\bibfield{author}{\bibinfo{person}{Xiangnan He}, \bibinfo{person}{Lizi Liao},
  \bibinfo{person}{Hanwang Zhang}, \bibinfo{person}{Liqiang Nie},
  \bibinfo{person}{Xia Hu}, {and} \bibinfo{person}{Tat-Seng Chua}.}
  \bibinfo{year}{2017}\natexlab{}.
\newblock \showarticletitle{Neural collaborative filtering}. In
  \bibinfo{booktitle}{\emph{WWW}}. \bibinfo{pages}{173--182}.
\newblock


\bibitem[\protect\citeauthoryear{Hsieh, Yang, Cui, Lin, Belongie, and
  Estrin}{Hsieh et~al\mbox{.}}{2017}]%
        {CML2017}
\bibfield{author}{\bibinfo{person}{Cheng-Kang Hsieh}, \bibinfo{person}{Longqi
  Yang}, \bibinfo{person}{Yin Cui}, \bibinfo{person}{Tsung-Yi Lin},
  \bibinfo{person}{Serge Belongie}, {and} \bibinfo{person}{Deborah Estrin}.}
  \bibinfo{year}{2017}\natexlab{}.
\newblock \showarticletitle{Collaborative metric learning}. In
  \bibinfo{booktitle}{\emph{WWW}}. \bibinfo{pages}{193--201}.
\newblock


\bibitem[\protect\citeauthoryear{Hu, Koren, and Volinsky}{Hu
  et~al\mbox{.}}{2008}]%
        {wrmf2008}
\bibfield{author}{\bibinfo{person}{Yifan Hu}, \bibinfo{person}{Yehuda Koren},
  {and} \bibinfo{person}{Chris Volinsky}.} \bibinfo{year}{2008}\natexlab{}.
\newblock \showarticletitle{Collaborative filtering for implicit feedback
  datasets}. In \bibinfo{booktitle}{\emph{ICDM}}. Ieee,
  \bibinfo{pages}{263--272}.
\newblock


\bibitem[\protect\citeauthoryear{Huang, Dong, Ding, Yang, Feng, Wang, and
  Tang}{Huang et~al\mbox{.}}{2021}]%
        {huang2021mixgcf}
\bibfield{author}{\bibinfo{person}{Tinglin Huang}, \bibinfo{person}{Yuxiao
  Dong}, \bibinfo{person}{Ming Ding}, \bibinfo{person}{Zhen Yang},
  \bibinfo{person}{Wenzheng Feng}, \bibinfo{person}{Xinyu Wang}, {and}
  \bibinfo{person}{Jie Tang}.} \bibinfo{year}{2021}\natexlab{}.
\newblock \showarticletitle{MixGCF: An Improved Training Method for Graph
  Neural Network-based Recommender Systems}. In
  \bibinfo{booktitle}{\emph{KDD}}. \bibinfo{pages}{665--674}.
\newblock


\bibitem[\protect\citeauthoryear{Kipf and Welling}{Kipf and Welling}{2017}]%
        {gcn2017}
\bibfield{author}{\bibinfo{person}{Thomas~N Kipf} {and} \bibinfo{person}{Max
  Welling}.} \bibinfo{year}{2017}\natexlab{}.
\newblock \showarticletitle{Semi-Supervised Classification with Graph
  Convolutional Networks}. In \bibinfo{booktitle}{\emph{ICLR}}.
\newblock


\bibitem[\protect\citeauthoryear{Koren}{Koren}{2008}]%
        {koren2008factorization}
\bibfield{author}{\bibinfo{person}{Yehuda Koren}.}
  \bibinfo{year}{2008}\natexlab{}.
\newblock \showarticletitle{Factorization meets the neighborhood: a
  multifaceted collaborative filtering model}. In
  \bibinfo{booktitle}{\emph{KDD}}. \bibinfo{pages}{426--434}.
\newblock


\bibitem[\protect\citeauthoryear{Koren, Bell, and Volinsky}{Koren
  et~al\mbox{.}}{2009}]%
        {koren2009matrix}
\bibfield{author}{\bibinfo{person}{Yehuda Koren}, \bibinfo{person}{Robert
  Bell}, {and} \bibinfo{person}{Chris Volinsky}.}
  \bibinfo{year}{2009}\natexlab{}.
\newblock \showarticletitle{Matrix factorization techniques for recommender
  systems}.
\newblock \bibinfo{journal}{\emph{Computer}} \bibinfo{volume}{42},
  \bibinfo{number}{8} (\bibinfo{year}{2009}), \bibinfo{pages}{30--37}.
\newblock


\bibitem[\protect\citeauthoryear{Li, Zhou, Zhang, Yang, Lian, and Huang}{Li
  et~al\mbox{.}}{2022b}]%
        {li2022bsal}
\bibfield{author}{\bibinfo{person}{Bisheng Li}, \bibinfo{person}{Min Zhou},
  \bibinfo{person}{Shengzhong Zhang}, \bibinfo{person}{Menglin Yang},
  \bibinfo{person}{Defu Lian}, {and} \bibinfo{person}{Zengfeng Huang}.}
  \bibinfo{year}{2022}\natexlab{b}.
\newblock \showarticletitle{BSAL: A Framework of Bi-Component Structure and
  Attribute Learning for Link Prediction}. In
  \bibinfo{booktitle}{\emph{SIGIR}}.
\newblock


\bibitem[\protect\citeauthoryear{Li, Gao, Bing, King, and Lyu}{Li
  et~al\mbox{.}}{2019}]%
        {li2019improving}
\bibfield{author}{\bibinfo{person}{Jingjing Li}, \bibinfo{person}{Yifan Gao},
  \bibinfo{person}{Lidong Bing}, \bibinfo{person}{Irwin King}, {and}
  \bibinfo{person}{Michael~R Lyu}.} \bibinfo{year}{2019}\natexlab{}.
\newblock \showarticletitle{Improving question generation with to the point
  context}.
\newblock \bibinfo{journal}{\emph{arXiv preprint arXiv:1910.06036}}
  (\bibinfo{year}{2019}).
\newblock


\bibitem[\protect\citeauthoryear{Li, Li, Ge, King, and Lyu}{Li
  et~al\mbox{.}}{2022a}]%
        {li2022text}
\bibfield{author}{\bibinfo{person}{Jingjing Li}, \bibinfo{person}{Zichao Li},
  \bibinfo{person}{Tao Ge}, \bibinfo{person}{Irwin King}, {and}
  \bibinfo{person}{Michael~R Lyu}.} \bibinfo{year}{2022}\natexlab{a}.
\newblock \showarticletitle{Text Revision by On-the-Fly Representation
  Optimization}.
\newblock \bibinfo{journal}{\emph{arXiv preprint arXiv:2204.07359}}
  (\bibinfo{year}{2022}).
\newblock


\bibitem[\protect\citeauthoryear{Li, Li, Mou, Jiang, Lyu, and King}{Li
  et~al\mbox{.}}{2020a}]%
        {li2020unsupervised}
\bibfield{author}{\bibinfo{person}{Jingjing Li}, \bibinfo{person}{Zichao Li},
  \bibinfo{person}{Lili Mou}, \bibinfo{person}{Xin Jiang},
  \bibinfo{person}{Michael Lyu}, {and} \bibinfo{person}{Irwin King}.}
  \bibinfo{year}{2020}\natexlab{a}.
\newblock \showarticletitle{Unsupervised text generation by learning from
  search}. In \bibinfo{booktitle}{\emph{NeurIPS}}, Vol.~\bibinfo{volume}{33}.
  \bibinfo{pages}{10820--10831}.
\newblock


\bibitem[\protect\citeauthoryear{Li, Zhang, Zhu, Qian, Zang, Han, and Hu}{Li
  et~al\mbox{.}}{2020b}]%
        {SML2020}
\bibfield{author}{\bibinfo{person}{Mingming Li}, \bibinfo{person}{Shuai Zhang},
  \bibinfo{person}{Fuqing Zhu}, \bibinfo{person}{Wanhui Qian},
  \bibinfo{person}{Liangjun Zang}, \bibinfo{person}{Jizhong Han}, {and}
  \bibinfo{person}{Songlin Hu}.} \bibinfo{year}{2020}\natexlab{b}.
\newblock \showarticletitle{Symmetric metric learning with adaptive margin for
  recommendation}. In \bibinfo{booktitle}{\emph{AAAI}},
  Vol.~\bibinfo{volume}{34}. \bibinfo{pages}{4634--4641}.
\newblock


\bibitem[\protect\citeauthoryear{Liang, Krishnan, Hoffman, and Jebara}{Liang
  et~al\mbox{.}}{2018}]%
        {VAECF2018}
\bibfield{author}{\bibinfo{person}{Dawen Liang}, \bibinfo{person}{Rahul~G
  Krishnan}, \bibinfo{person}{Matthew~D Hoffman}, {and} \bibinfo{person}{Tony
  Jebara}.} \bibinfo{year}{2018}\natexlab{}.
\newblock \showarticletitle{Variational autoencoders for collaborative
  filtering}. In \bibinfo{booktitle}{\emph{WWW}}. \bibinfo{pages}{689--698}.
\newblock


\bibitem[\protect\citeauthoryear{Liu, Yang, Zhou, Feng, and Fournier-Viger}{Liu
  et~al\mbox{.}}{2022}]%
        {liu2022enhancing}
\bibfield{author}{\bibinfo{person}{Jiahong Liu}, \bibinfo{person}{Menglin
  Yang}, \bibinfo{person}{Min Zhou}, \bibinfo{person}{Shanshan Feng}, {and}
  \bibinfo{person}{Philippe Fournier-Viger}.} \bibinfo{year}{2022}\natexlab{}.
\newblock \showarticletitle{Enhancing Hyperbolic Graph Embeddings via
  Contrastive Learning}.
\newblock \bibinfo{journal}{\emph{arXiv preprint arXiv:2201.08554}}
  (\bibinfo{year}{2022}).
\newblock


\bibitem[\protect\citeauthoryear{Liu, Nickel, and Kiela}{Liu
  et~al\mbox{.}}{2019}]%
        {liu2019HGNN}
\bibfield{author}{\bibinfo{person}{Qi Liu}, \bibinfo{person}{Maximilian
  Nickel}, {and} \bibinfo{person}{Douwe Kiela}.}
  \bibinfo{year}{2019}\natexlab{}.
\newblock \showarticletitle{Hyperbolic graph neural networks}. In
  \bibinfo{booktitle}{\emph{NeurIPS}}. \bibinfo{pages}{8230--8241}.
\newblock


\bibitem[\protect\citeauthoryear{Luo, Zhou, Xia, and Zhu}{Luo
  et~al\mbox{.}}{2014}]%
        {nmf-cf}
\bibfield{author}{\bibinfo{person}{Xin Luo}, \bibinfo{person}{Mengchu Zhou},
  \bibinfo{person}{Yunni Xia}, {and} \bibinfo{person}{Qingsheng Zhu}.}
  \bibinfo{year}{2014}\natexlab{}.
\newblock \showarticletitle{An efficient non-negative
  matrix-factorization-based approach to collaborative filtering for
  recommender systems}.
\newblock \bibinfo{journal}{\emph{TII}} \bibinfo{volume}{10},
  \bibinfo{number}{2} (\bibinfo{year}{2014}), \bibinfo{pages}{1273--1284}.
\newblock


\bibitem[\protect\citeauthoryear{Mao, Zhu, Wang, Dai, Dong, Xiao, and He}{Mao
  et~al\mbox{.}}{2021a}]%
        {mao2021simplex}
\bibfield{author}{\bibinfo{person}{Kelong Mao}, \bibinfo{person}{Jieming Zhu},
  \bibinfo{person}{Jinpeng Wang}, \bibinfo{person}{Quanyu Dai},
  \bibinfo{person}{Zhenhua Dong}, \bibinfo{person}{Xi Xiao}, {and}
  \bibinfo{person}{Xiuqiang He}.} \bibinfo{year}{2021}\natexlab{a}.
\newblock \showarticletitle{SimpleX: A Simple and Strong Baseline for
  Collaborative Filtering}. In \bibinfo{booktitle}{\emph{CIKM}}.
  \bibinfo{pages}{1243--1252}.
\newblock


\bibitem[\protect\citeauthoryear{Mao, Zhu, Xiao, Lu, Wang, and He}{Mao
  et~al\mbox{.}}{2021b}]%
        {mao2021ultragcn}
\bibfield{author}{\bibinfo{person}{Kelong Mao}, \bibinfo{person}{Jieming Zhu},
  \bibinfo{person}{Xi Xiao}, \bibinfo{person}{Biao Lu},
  \bibinfo{person}{Zhaowei Wang}, {and} \bibinfo{person}{Xiuqiang He}.}
  \bibinfo{year}{2021}\natexlab{b}.
\newblock \showarticletitle{UltraGCN: ultra simplification of graph
  convolutional networks for recommendation}. In
  \bibinfo{booktitle}{\emph{CIKM}}. \bibinfo{pages}{1253--1262}.
\newblock


\bibitem[\protect\citeauthoryear{Nickel and Kiela}{Nickel and Kiela}{2017}]%
        {nickel2017poincare}
\bibfield{author}{\bibinfo{person}{Maximillian Nickel} {and}
  \bibinfo{person}{Douwe Kiela}.} \bibinfo{year}{2017}\natexlab{}.
\newblock \showarticletitle{Poincar{\'e} embeddings for learning hierarchical
  representations}. In \bibinfo{booktitle}{\emph{NeurIPS}}.
  \bibinfo{pages}{6338--6347}.
\newblock


\bibitem[\protect\citeauthoryear{Nickel and Kiela}{Nickel and Kiela}{2018}]%
        {nickel2018learning}
\bibfield{author}{\bibinfo{person}{Maximillian Nickel} {and}
  \bibinfo{person}{Douwe Kiela}.} \bibinfo{year}{2018}\natexlab{}.
\newblock \showarticletitle{Learning Continuous Hierarchies in the Lorentz
  Model of Hyperbolic Geometry}. In \bibinfo{booktitle}{\emph{ICML}}.
  \bibinfo{pages}{3779--3788}.
\newblock


\bibitem[\protect\citeauthoryear{Park, Kim, Xie, and Yu}{Park
  et~al\mbox{.}}{2018}]%
        {park2018collaborative}
\bibfield{author}{\bibinfo{person}{Chanyoung Park}, \bibinfo{person}{Donghyun
  Kim}, \bibinfo{person}{Xing Xie}, {and} \bibinfo{person}{Hwanjo Yu}.}
  \bibinfo{year}{2018}\natexlab{}.
\newblock \showarticletitle{Collaborative translational metric learning}. In
  \bibinfo{booktitle}{\emph{ICDM}}. IEEE, \bibinfo{pages}{367--376}.
\newblock


\bibitem[\protect\citeauthoryear{Sala, De~Sa, Gu, and Re}{Sala
  et~al\mbox{.}}{2018}]%
        {sala2018representation}
\bibfield{author}{\bibinfo{person}{Frederic Sala}, \bibinfo{person}{Chris
  De~Sa}, \bibinfo{person}{Albert Gu}, {and} \bibinfo{person}{Christopher Re}.}
  \bibinfo{year}{2018}\natexlab{}.
\newblock \showarticletitle{Representation Tradeoffs for Hyperbolic
  Embeddings}. In \bibinfo{booktitle}{\emph{ICML}}.
  \bibinfo{pages}{4460--4469}.
\newblock


\bibitem[\protect\citeauthoryear{Song, Meng, Zhang, and King}{Song
  et~al\mbox{.}}{2021}]%
        {zixingcikm2021}
\bibfield{author}{\bibinfo{person}{Zixing Song}, \bibinfo{person}{Ziqiao Meng},
  \bibinfo{person}{Yifei Zhang}, {and} \bibinfo{person}{Irwin King}.}
  \bibinfo{year}{2021}\natexlab{}.
\newblock \showarticletitle{Semi-supervised Multi-label Learning for
  Graph-structured Data}. In \bibinfo{booktitle}{\emph{{CIKM}}}.
  \bibinfo{publisher}{{ACM}}, \bibinfo{pages}{1723--1733}.
\newblock


\bibitem[\protect\citeauthoryear{Song, Yang, Xu, and King}{Song
  et~al\mbox{.}}{2022}]%
        {zixingsurvey}
\bibfield{author}{\bibinfo{person}{Zixing Song}, \bibinfo{person}{Xiangli
  Yang}, \bibinfo{person}{Zenglin Xu}, {and} \bibinfo{person}{Irwin King}.}
  \bibinfo{year}{2022}\natexlab{}.
\newblock \showarticletitle{Graph-Based Semi-Supervised Learning: A
  Comprehensive Review}.
\newblock \bibinfo{journal}{\emph{TNNLS}} (\bibinfo{year}{2022}),
  \bibinfo{pages}{1--21}.
\newblock


\bibitem[\protect\citeauthoryear{Sun, Cheng, Zuberi, P{\'e}rez, and
  Volkovs}{Sun et~al\mbox{.}}{2021}]%
        {sun2021hgcf}
\bibfield{author}{\bibinfo{person}{Jianing Sun}, \bibinfo{person}{Zhaoyue
  Cheng}, \bibinfo{person}{Saba Zuberi}, \bibinfo{person}{Felipe P{\'e}rez},
  {and} \bibinfo{person}{Maksims Volkovs}.} \bibinfo{year}{2021}\natexlab{}.
\newblock \showarticletitle{HGCF: Hyperbolic Graph Convolution Networks for
  Collaborative Filtering}. In \bibinfo{booktitle}{\emph{WWW}}.
  \bibinfo{pages}{593--601}.
\newblock


\bibitem[\protect\citeauthoryear{Tay, Anh~Tuan, and Hui}{Tay
  et~al\mbox{.}}{2018}]%
        {LRML2018}
\bibfield{author}{\bibinfo{person}{Yi Tay}, \bibinfo{person}{Luu Anh~Tuan},
  {and} \bibinfo{person}{Siu~Cheung Hui}.} \bibinfo{year}{2018}\natexlab{}.
\newblock \showarticletitle{Latent relational metric learning via memory-based
  attention for collaborative ranking}. In \bibinfo{booktitle}{\emph{WWW}}.
  \bibinfo{pages}{729--739}.
\newblock


\bibitem[\protect\citeauthoryear{Veli{\v{c}}kovi{\'c}, Cucurull, Casanova,
  Romero, Lio, and Bengio}{Veli{\v{c}}kovi{\'c} et~al\mbox{.}}{2017}]%
        {GAT}
\bibfield{author}{\bibinfo{person}{Petar Veli{\v{c}}kovi{\'c}},
  \bibinfo{person}{Guillem Cucurull}, \bibinfo{person}{Arantxa Casanova},
  \bibinfo{person}{Adriana Romero}, \bibinfo{person}{Pietro Lio}, {and}
  \bibinfo{person}{Yoshua Bengio}.} \bibinfo{year}{2017}\natexlab{}.
\newblock \showarticletitle{Graph attention networks}.
\newblock \bibinfo{journal}{\emph{arXiv preprint arXiv:1710.10903}}
  (\bibinfo{year}{2017}).
\newblock


\bibitem[\protect\citeauthoryear{Vinh~Tran, Tay, Zhang, Cong, and Li}{Vinh~Tran
  et~al\mbox{.}}{2020}]%
        {HyperML2020}
\bibfield{author}{\bibinfo{person}{Lucas Vinh~Tran}, \bibinfo{person}{Yi Tay},
  \bibinfo{person}{Shuai Zhang}, \bibinfo{person}{Gao Cong}, {and}
  \bibinfo{person}{Xiaoli Li}.} \bibinfo{year}{2020}\natexlab{}.
\newblock \showarticletitle{Hyper{ML}: A Boosting Metric Learning Approach in
  Hyperbolic Space for Recommender Systems}. In
  \bibinfo{booktitle}{\emph{WSDM}}. \bibinfo{address}{New York, NY, USA},
  \bibinfo{pages}{609–617}.
\newblock


\bibitem[\protect\citeauthoryear{Wang, Lian, Tong, Liu, Huang, and Chen}{Wang
  et~al\mbox{.}}{2021}]%
        {wang2021hypersorec}
\bibfield{author}{\bibinfo{person}{Hao Wang}, \bibinfo{person}{Defu Lian},
  \bibinfo{person}{Hanghang Tong}, \bibinfo{person}{Qi Liu},
  \bibinfo{person}{Zhenya Huang}, {and} \bibinfo{person}{Enhong Chen}.}
  \bibinfo{year}{2021}\natexlab{}.
\newblock \showarticletitle{HyperSoRec: Exploiting Hyperbolic User and Item
  Representations with Multiple Aspects for Social-aware Recommendation}.
\newblock \bibinfo{journal}{\emph{TOIS}} (\bibinfo{year}{2021}),
  \bibinfo{pages}{1–28}.
\newblock


\bibitem[\protect\citeauthoryear{Wang, He, Wang, Feng, and Chua}{Wang
  et~al\mbox{.}}{2019}]%
        {wang2019ngcf}
\bibfield{author}{\bibinfo{person}{Xiang Wang}, \bibinfo{person}{Xiangnan He},
  \bibinfo{person}{Meng Wang}, \bibinfo{person}{Fuli Feng}, {and}
  \bibinfo{person}{Tat-Seng Chua}.} \bibinfo{year}{2019}\natexlab{}.
\newblock \showarticletitle{Neural graph collaborative filtering}. In
  \bibinfo{booktitle}{\emph{SIGIR}}. \bibinfo{pages}{165--174}.
\newblock


\bibitem[\protect\citeauthoryear{Yang, Chen, Li, Yu, and Xu}{Yang
  et~al\mbox{.}}{2021a}]%
        {yang2021hyper}
\bibfield{author}{\bibinfo{person}{Haoran Yang}, \bibinfo{person}{Hongxu Chen},
  \bibinfo{person}{Lin Li}, \bibinfo{person}{Philip~S Yu}, {and}
  \bibinfo{person}{Guandong Xu}.} \bibinfo{year}{2021}\natexlab{a}.
\newblock \showarticletitle{Hyper Meta-Path Contrastive Learning for
  Multi-Behavior Recommendation}.
\newblock \bibinfo{journal}{\emph{arXiv preprint arXiv:2109.02859}}
  (\bibinfo{year}{2021}).
\newblock


\bibitem[\protect\citeauthoryear{Yang, Meng, and King}{Yang
  et~al\mbox{.}}{2020}]%
        {FeatureNorm2020}
\bibfield{author}{\bibinfo{person}{Menglin Yang}, \bibinfo{person}{Ziqiao
  Meng}, {and} \bibinfo{person}{Irwin King}.} \bibinfo{year}{2020}\natexlab{}.
\newblock \showarticletitle{FeatureNorm: L2 Feature Normalization for Dynamic
  Graph Embedding}. In \bibinfo{booktitle}{\emph{ICDM}}.
  \bibinfo{pages}{731--740}.
\newblock


\bibitem[\protect\citeauthoryear{Yang, Zhou, Kalander, Huang, and King}{Yang
  et~al\mbox{.}}{2021b}]%
        {yang2021discrete}
\bibfield{author}{\bibinfo{person}{Menglin Yang}, \bibinfo{person}{Min Zhou},
  \bibinfo{person}{Marcus Kalander}, \bibinfo{person}{Zengfeng Huang}, {and}
  \bibinfo{person}{Irwin King}.} \bibinfo{year}{2021}\natexlab{b}.
\newblock \showarticletitle{Discrete-time Temporal Network Embedding via
  Implicit Hierarchical Learning in Hyperbolic Space}. In
  \bibinfo{booktitle}{\emph{KDD}}. \bibinfo{pages}{1975--1985}.
\newblock


\bibitem[\protect\citeauthoryear{Yang, Zhou, Li, Liu, Pan, Xiong, and
  King}{Yang et~al\mbox{.}}{2022a}]%
        {yang2022hyperbolic}
\bibfield{author}{\bibinfo{person}{Menglin Yang}, \bibinfo{person}{Min Zhou},
  \bibinfo{person}{Zhihao Li}, \bibinfo{person}{Jiahong Liu},
  \bibinfo{person}{Lujia Pan}, \bibinfo{person}{Hui Xiong}, {and}
  \bibinfo{person}{Irwin King}.} \bibinfo{year}{2022}\natexlab{a}.
\newblock \showarticletitle{Hyperbolic Graph Neural Networks: A Review of
  Methods and Applications}.
\newblock \bibinfo{journal}{\emph{arXiv preprint arXiv:2202.13852}}
  (\bibinfo{year}{2022}).
\newblock


\bibitem[\protect\citeauthoryear{Yang, Zhou, Liu, Lian, and King}{Yang
  et~al\mbox{.}}{2022b}]%
        {yang2022hrcf}
\bibfield{author}{\bibinfo{person}{Menglin Yang}, \bibinfo{person}{Min Zhou},
  \bibinfo{person}{Jiahong Liu}, \bibinfo{person}{Defu Lian}, {and}
  \bibinfo{person}{Irwin King}.} \bibinfo{year}{2022}\natexlab{b}.
\newblock \showarticletitle{HRCF: Enhancing collaborative filtering via
  hyperbolic geometric regularization}. In \bibinfo{booktitle}{\emph{WWW}}.
  \bibinfo{pages}{2462--2471}.
\newblock


\bibitem[\protect\citeauthoryear{Ying, He, Chen, Eksombatchai, Hamilton, and
  Leskovec}{Ying et~al\mbox{.}}{2018}]%
        {ying2018graph}
\bibfield{author}{\bibinfo{person}{Rex Ying}, \bibinfo{person}{Ruining He},
  \bibinfo{person}{Kaifeng Chen}, \bibinfo{person}{Pong Eksombatchai},
  \bibinfo{person}{William~L Hamilton}, {and} \bibinfo{person}{Jure Leskovec}.}
  \bibinfo{year}{2018}\natexlab{}.
\newblock \showarticletitle{Graph convolutional neural networks for web-scale
  recommender systems}. In \bibinfo{booktitle}{\emph{KDD}}.
  \bibinfo{pages}{974--983}.
\newblock


\bibitem[\protect\citeauthoryear{Zhang, Chen, Ming, Cui, Yin, and Xu}{Zhang
  et~al\mbox{.}}{2021a}]%
        {zhang2021we}
\bibfield{author}{\bibinfo{person}{Sixiao Zhang}, \bibinfo{person}{Hongxu
  Chen}, \bibinfo{person}{Xiao Ming}, \bibinfo{person}{Lizhen Cui},
  \bibinfo{person}{Hongzhi Yin}, {and} \bibinfo{person}{Guandong Xu}.}
  \bibinfo{year}{2021}\natexlab{a}.
\newblock \showarticletitle{Where are we in embedding spaces? A Comprehensive
  Analysis on Network Embedding Approaches for Recommender Systems}. In
  \bibinfo{booktitle}{\emph{KDD}}.
\newblock


\bibitem[\protect\citeauthoryear{Zhang, Wang, Shi, Liu, and Song}{Zhang
  et~al\mbox{.}}{2021b}]%
        {lgcn}
\bibfield{author}{\bibinfo{person}{Yiding Zhang}, \bibinfo{person}{Xiao Wang},
  \bibinfo{person}{Chuan Shi}, \bibinfo{person}{Nian Liu}, {and}
  \bibinfo{person}{Guojie Song}.} \bibinfo{year}{2021}\natexlab{b}.
\newblock \showarticletitle{Lorentzian Graph Convolutional Networks}. In
  \bibinfo{booktitle}{\emph{WWW}}. \bibinfo{pages}{1249--1261}.
\newblock


\bibitem[\protect\citeauthoryear{Zhang, Zhu, Meng, Koniusz, and King}{Zhang
  et~al\mbox{.}}{2022}]%
        {zhang2022graph}
\bibfield{author}{\bibinfo{person}{Yifei Zhang}, \bibinfo{person}{Hao Zhu},
  \bibinfo{person}{Ziqiao Meng}, \bibinfo{person}{Piotr Koniusz}, {and}
  \bibinfo{person}{Irwin King}.} \bibinfo{year}{2022}\natexlab{}.
\newblock \showarticletitle{Graph-adaptive rectified linear unit for graph
  neural networks}. In \bibinfo{booktitle}{\emph{WWW}}.
  \bibinfo{pages}{1331--1339}.
\newblock


\bibitem[\protect\citeauthoryear{Zhou, Li, Yang, and Pan}{Zhou
  et~al\mbox{.}}{2022}]%
        {zhou2022telegraph}
\bibfield{author}{\bibinfo{person}{Min Zhou}, \bibinfo{person}{Bisheng Li},
  \bibinfo{person}{Menglin Yang}, {and} \bibinfo{person}{Lujia Pan}.}
  \bibinfo{year}{2022}\natexlab{}.
\newblock \showarticletitle{TeleGraph: A Benchmark Dataset for Hierarchical
  Link Prediction}.
\newblock \bibinfo{journal}{\emph{arXiv preprint arXiv:2204.07703}}
  (\bibinfo{year}{2022}).
\newblock


\end{thebibliography}
\newpage
\appendix
\onecolumn
\section{Comparison with Other Negative Samplings}
To our knowledge, there is no technique, especially for performing negative sampling in hyperbolic recommender systems. In this section, we compare three widely used techniques to obtain negative sampling in Euclidean models: dynamic random sampling (DS), popularity-based sampling (Popularity) and mix feature sampling (Mix).
Dynamic random sampling is used to randomly select a node as the negative item in each iteration.
The purpose of popularity-based sampling is to select negative items based on their degrees, which could reflect their popularity. The sampling process is performed before the training process, which is a static method.
The mixing feature method is based on the idea of~\cite{huang2021mixgcf}, that is, creating a negative sample based on the mix of positive and negative items. Mixing method obtains negative samples in each iteration which is a dynamic method.

As shown in Table~\ref{tab:negative_sampling}, we found that among the three sampling methods, DS sampling can improve the performance of recommendations to some extent. The reason lies that compared to static methods, dynamic methods help the model learn more negative samples and obtain enough information. However, DS is insufficient, as most negative samples are trivial or noisy. Combined with the performance of HICF (ours), we know that the proposed HINS is much more helpful. In addition, the other two methods based on popularity and fusion methods are not compatible with hyperbolic models, which perform much worse on hyperbolic models.
\begin{table*}[!t]
\centering
\caption{Comparisons with different negative sampling techniques.}
\label{tab:negative_sampling}
\resizebox{0.94\textwidth}{!}{%
\begin{tabular}{@{}l|cccc|cccc|cccc@{}}
\toprule
Dataset   & \multicolumn{4}{c|}{Amazon-CD}      & \multicolumn{4}{c|}{Amazon-Book}    & \multicolumn{4}{c}{Yelp2020}           \\ 
Metrics   & DS     & Popularity & Mix    & Ours   & DS     & Popularity & Mix    & Ours   & DS     & Popularity & Mix    & Ours   \\ \midrule
Recall@10 & 0.1013 & 0.0912  & 0.0884 & \textbf{0.1079} & 0.0858 & 0.0860  & 0.0757 & \textbf{0.0965} & 0.0542 & 0.0528  & 0.0454 & \textbf{0.0570}  \\
NDCG@10   & 0.0793 & 0.0708  & 0.0684 & \textbf{0.0848} & 0.0842 & 0.0860  & 0.0749 & \textbf{0.0978} & 0.0470 & 0.0460  & 0.0396 & \textbf{0.0502} \\
Recall@20 & 0.1504 & 0.1338  & 0.1362 & \textbf{0.0965} & 0.1310 & 0.1317  & 0.1175 & \textbf{0.1449} & 0.0898 & 0.0890  & 0.0776 & \textbf{0.0948} \\
NDCG@20   & 0.0953 & 0.0846  & 0.0839 & \textbf{0.1010} & 0.1000 & 0.1018  & 0.0894 & \textbf{0.1142} & 0.0595 & 0.0587  & 0.0508 & \textbf{0.0633} \\ \bottomrule
\end{tabular}%
}
\end{table*}

\end{document}